\documentclass{aa}
\usepackage{latexsym}
\usepackage{epsfig}
\usepackage{graphicx}
\newcommand{\radm}{rad~m$^{-2}$} 
\newcommand{\pl}{\parallel}
\newcommand{\dg}{\degr}
\newcommand{\srm}{\mbox{$\sigma_{\scriptscriptstyle\rm RM}$}}
\newcommand{\HI}{H {\sc i}\ }
\renewcommand{\ga}{\,\hbox{\rlap{\hbox{\lower4pt\hbox{$\sim$}}}
                  \lower-1pt\hbox{\hspace{-4pt}$>$}}\,}
\begin{document}

\title{Multi-frequency polarimetry of the Galactic radio background
       around 350~MHz: II. A region in Horologium around \\l = 137\dg,
       b = 7\dg}

\author{M. Haverkorn\inst{1}
        \and
	P. Katgert\inst{2}
        \and
	A. G. de Bruyn\inst{3,4}
        }
\offprints{M. Haverkorn}
\institute{Leiden Observatory, P.O.Box 9513, 2300 RA Leiden, the
           Netherlands\\
	   (Current address: Harvard-Smithsonian Center for
	   Astrophysics, 60 Garden Street MS-67, Cambridge MA
	   02138, USA)\\
           \email{mhaverkorn@cfa.harvard.edu} 
      \and Leiden Observatory, P.O.Box 9513, 2300 RA Leiden, the
           Netherlands\\ 
           \email{katgert@strw.leidenuniv.nl} 
      \and ASTRON, P.O.Box 2, 7990 AA Dwingeloo, the Netherlands\\ 
           \email{ger@astron.nl}
      \and Kapteyn Institute, P.O.Box 800, 9700 AV Groningen,
           the Netherlands }
\titlerunning{Multi-frequency polarimetry around 350~MHz II}

\abstract{We study a conspicuous ring-like structure with a radius of
   about  1.4\dg\ which was observed with the Westerbork Synthesis
   Radio Telescope (WSRT) at 5 frequencies around 350~MHz. This ring
   is very prominent in Stokes $Q$ and $U$, and less so in polarized
   intensity~$P$. No corresponding structure is visible in total
   intensity Stokes~$I$, which indicates that the ring is created by
   Faraday rotation and depolarization processes. The polarization
   angle changes from the center of the ring outwards to a radius $\ga
   1.7\dg$. Thus, the structure in polarization angle is not ring-like
   but resembles a disk, and it is larger than the ring in~$P$. The
   rotation measure $RM$ decreases almost continuously over the disk,
   from $RM$~$\approx 0$~\radm\ at the edge, to $-8$~\radm\ in the
   center, while outside the ring the $RM$ is slightly positive. This
   radial variation of $RM$ yields stringent constraints on the nature
   of the ring-like structure, because it rules out any spherically
   symmetrical magnetic field configuration, such as might be expected
   from supernova remnants or wind-blown bubbles.  We discuss several
   possible connections between the ring and known objects in the ISM,
   and conclude that the ring is a predominantly magnetic funnel-like
   structure. This description can explain both the field reversal
   from outside to inside the ring, and the increase in magnetic
   field, probably combined with an electron density increase, towards
   the center of the ring. The ring-structure in~$P$ is most likely
   caused by a lack of depolarization due to a very uniform $RM$
   distribution at that radius.  Beyond the ring, the $RM$ gradient
   increases, depolarizing the polarized emission, so that the
   polarized intensity decreases. In the southwestern corner of the
   field a pattern of narrow filaments of low polarization, aligned
   with Galactic longitude, is observed, indicative of beam
   depolarization due to abrupt changes in $RM$. This explanation is
   supported by the observed $RM$.
    }
   
\maketitle

\section{Introduction}
\label{s5:intro}

Observation of the diffuse polarized background of our Galaxy at radio
frequencies provides a unique way to observe features in the ISM that
are otherwise invisible. In this paper we report and analyze the
detection in radio polarization of a remarkable circular structure
with a radius of $\sim 1.4\dg$. The observed radiation is synchrotron
radiation, originating from relativistic cosmic ray electrons in
interaction with the Galactic magnetic field. The linearly polarized
component of the synchrotron radiation is modulated by several
mechanisms, among which Faraday rotation, viz.\ the birefringence of
left- and right-handed circularly polarized radiation in a medium
which contains a magnetic field and free electrons. For linear
polarization, this results in a rotation of the angle of polarization
$\phi$ in passage through a magneto-ionized medium. This rotation is
proportional to the square of the wavelength where the proportionality
constant is the rotation measure ($RM$). The $RM$ depends on the
magnetic field component parallel to the line of sight $B_{\pl}$,
weighted by the thermal electron density $n_e$, integrated over the
line of sight. Thus multi-wavelength observations of linear
polarization yield directly the electron-density-weighted value of the
interstellar magnetic field.

The circular structure we have observed at $(l,b) \approx (137\dg,
7\dg)$ in the constellation Horologium is located in the middle of a
region of very high polarization extending over many degrees as
described by e.g.\ Brouw \& Spoelstra (1976).

Bingham \& Shakeshaft (1967) were the first to observe the circular
structure in this region in a map of $RM$ that they constructed from
surveys by Berkhuijsen et al.\ (1963, 1964) at 408 MHz and 610 MHz, by
Wielebinski \& Shakeshaft (1964) at 408 MHz and by Bingham (1966) at
1407 MHz. The $RM$ map of Bingham \& Shakeshaft shows a circular
structure of about the size of their beam ($\sim 2\dg$) where
$RM~\approx -5$~\radm, in a region where $RM$s are $-1$~\radm~$\la RM
\la 0$~\radm. Verschuur (1968) reobserved this region at higher
angular resolution ($\sim$~40\arcmin) with the 250ft MkI telescope at
Jodrell Bank at 408 MHz. Verschuur found that it is a ring-like
structure of {\em low} polarized intensity (50\% less than in its
surroundings) with a radius of a few degrees, and with a small region
(about one beam) of almost zero polarization at the southwestern edge
of the ring. Verschuur suggests that the region is connected to the
B2Ve star HD20336 which is located close to the center of the ring.

In a later paper, Verschuur (1969) presents \HI measurements obtained
with the 300ft Green Bank telescope which show an deficiency in \HI
along the trajectory of the star HD20336. He explains this by assuming
the star is moving through the neutral medium, expelling the \HI and
ionizing the remaining small fraction of neutral material. He does not
mention the ring-like structure.

The ring-like structure, which is mainly visible in Stokes $Q$ and $U$
and in polarization angle, was noticed in polarization maps produced
as a by-product of the Westerbork Northern Sky Survey (WENSS,
Rengelink et al.\ 1997, Schnitzeler et al.\ 2003).  In 1995/1996,
the region was reobserved with the Westerbork Synthesis Radio
Telescope (WSRT) at 8 frequencies by T. Spoelstra, who kindly allowed
us to analyze his observations.

In Sect.~\ref{s5:obs}, we discuss the observations, while in
Sect.~\ref{s5:res} the observational results are given.
Section~\ref{s5:prop} presents some observed properties of the
ring-like structure, that are subsequently interpreted in
Sect.~\ref{s5:or}. Possible connections of the ring structure to
other observed features in the Galaxy are presented in
Sect.~\ref{s5:conn}. In Sect.~\ref{s5:magn}, the ring is described as
a magnetic structure. In Sect.~\ref{s5:out}, we discuss remarkable
structure in the field unrelated to the ring, and Sect.~\ref{s5:conc}
presents our conclusions. 

\section{The observations}
\label{s5:obs}

The Westerbork Synthesis Radio Telescope (WSRT) was used to observe an
area of approximately $7\dg\times7\dg$, centered around $(l,b)
\approx (137\dg, 7\dg)$ in the constellation of Horologium. Data were
taken in 8 frequency bands simultaneously, each with a band width of 5~MHz,
but 3 frequency bands contain no usable data due to radio
interference. The 5 bands which contain good data are centered at
341~MHz, 349~MHz, 355~MHz, 360~MHz, and 375~MHz. The maximum
resolution of the WSRT array is $\sim$~1\arcmin, but the data were
smoothed using a Gaussian taper to obtain a
5.0\arcmin$\times$5.0\arcmin~cosec~$\delta =
5.0\arcmin\times5.5\arcmin$ resolution.  The derived maps of linearly
polarized intensities Stokes $Q$ and $U$ were used to compute the
polarized intensity $P$ and polarization angle $\phi$.  The noise in
Stokes $Q$ and $U$ was derived from the rms signal in (empty) Stokes
$V$ maps at 5.0\arcmin\ resolution and was $\sigma\approx$~5~mJy/beam
for all bands, see Table~\ref{t5:data}. The computation of the average
polarized intensity from $Q$ and $U$ is biased because $P$ is a
positive definite quantity. It can be debiased to first order as
$P_{debias} = \sqrt{(Q_{obs}^2 + U_{obs}^2) - \sigma^2}$ (for $P >
\sigma$). Here, $P$ generally has a S/N~$> 4-5$, for which the
debiasing does not alter the data by more than 2~-~3\%. Therefore, no
debiasing was applied.

To obtain a large field of view, and to reduce off-axis instrumental
polarization, the mosaicking technique was used. The telescope cycled
through a grid of pointing positions during the 12hr observation,
integrating 50 seconds per pointing, so that every pointing position
was observed many times per 12hr period. The mosaic was constructed
from $5\times5$ pointings, where the distance between the pointing
centers is 1.25\dg.  The edges of the mosaic far away from any pointing
center display high instrumental polarization, and these were not used
in the analysis.

The data were reduced with the {\sc newstar} reduction package, using
the unpolarized calibrator sources 3C48, 3C147 and 3C286, and the
polarized calibrators 3C345 and 3C303. The absolute flux scale at
325~MHz is based on a flux density of 26.93~Jy for 3C286 (Baars et
al.\ 1977). For details on the data reduction procedure, see Haverkorn
(2002).

\begin{table}
  \begin{center}
    \begin{tabular}{|l|ccc|}
      \hline
  Central position & \multicolumn{3}{l|}{(l,b) = (137\dg, 7\dg)} \\
  Size             & \multicolumn{3}{l|}{$\sim$~7\dg$\times$7\dg} \\        
  Pointings        & \multicolumn{3}{l|}{5$\times$5}                \\      
  Frequencies      & \multicolumn{3}{l|}{341, 349, 355, 360, 375 MHz} \\    
  Resolution       & 
  \multicolumn{3}{l|}{5.0\arcmin$\times$5.0\arcmin\ cosec $\delta$ 
    = 5.0\arcmin$\times$5.5\arcmin} \\                     
  Noise            & \multicolumn{3}{l|}{$\sim$~5~mJy/beam (0.7 K)}  \\     
  Conversion Jy-K  & \multicolumn{3}{l|}{1 mJy/beam = 0.146 K (at 350~MHz)}\\
  \hline\hline
  Minimal   & Date     & Start time & End time \\
  spacing   &          &  (UT)      &  (UT)\\
  \hline
  36m      & 95/12/19 & 15:12      & 02:58\\ 
  48m      & 95/12/20 & 14:53      & 02:53 \\ 
  60m      & 95/12/27 & 14:20      & 02:20 \\
  72m      & 96/01/02 & 14:02      & 02:02\\
  84m      & 95/12/12 & 16:42      & 03:25 \\
  96m      & 96/01/08 & 13:38      & 01:38 \\
  \hline  
    \end{tabular}
    \caption{Observational details of the Horologium field}
    \label{t5:data}
  \end{center}
\end{table}

To avoid radio interference from the Sun, the observations were taken
mostly or completely at night, as shown in Table~\ref{t5:data}.
Because the observations were done in winter, and during a solar
minimum, the total electron content (TEC) of the ionosphere is
low. These observing conditions minimize the ionospheric Faraday
rotation.  We can estimate the $RM$ contribution of the ionosphere
from the TEC of the ionosphere and the earth magnetic field. The TEC
of the ionosphere above Westerbork at night, in a solar minimum, and
in winter is minimal: TEC~$\approx 2.2 \; 10^{16}$~electrons~cm$^{-2}$
(Campbell, private communication).
Assuming a vertical component of the earth magnetic field of 4.5~G
down wards, and a path length through the ionosphere of 300~km, the
$RM$ caused by the ionosphere is --0.25~\radm\ at hour angle zero. At
larger hour angles, this is even less. So we expect the rotation
measure values given in this paper not to be affected by ionospheric
Faraday rotation by more than 0.5~\radm.

\subsection{Missing large scale structure}
\label{ss5:off}

An interferometer is insensitive to structure on large angular scales
due to missing short spacings. The smallest baseline attainable with
the WSRT of 36m means that scales above approximately a degree are
attenuated sufficiently as to be undetectable. The $Q$ and $U$ maps
are constructed so that in each mosaic pointing, the map integrals of
$Q$ and $U$ are zero. This leads to missing large-scale components in
$Q$ and $U$, and therefore erroneous determinations of $P$, $\phi$ and
$RM$. However, if the variation in $RM$ is large enough within one
pointing, the variation in polarization angle is so large that the
average $Q$ and $U$ are close to zero. In this case, missing
large-scale components are negligible. 

\begin{figure}
  \centering
  \psfig{figure=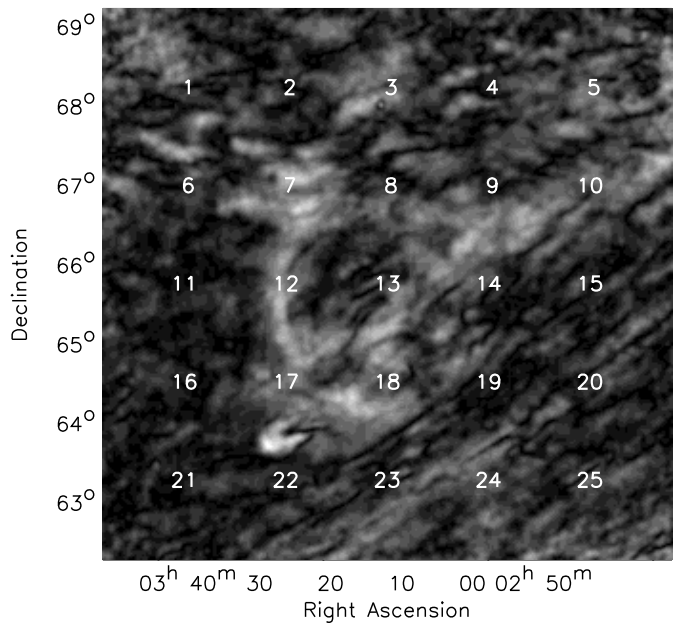,width=0.5\textwidth}
  \caption{Individual pointings are numbered 1 to 25, superimposed on
           polarized intensity at 349~MHz.}
  \label{f5:poi}
\end{figure}

In the field of observation
discussed here, \srm~$> 1.8$~\radm\ in most pointings, which means
that in these pointings missing large-scale components cannot account
to more than a few percent of the small-scale signal. The 25 pointing
centers are shown in Fig.~\ref{f5:poi}. In four pointings \srm~$\la
1.8$~\radm, which means that offsets could amount to 15 - 30\% of the
signal. These pointings are numbers 9, 10, 12 and 24, and the \srm\ in
those pointings are $\sim$~1.6, $\sim$~1.7, $\sim$~1.7 and~$\sim$~1.3
respectively. All \srm\ are computed only at positions
where $\langle P \rangle > 20$~mJy/beam and reduced $\chi^2$ of the
linear $\phi(\lambda^2)$-relation $\chi^2_{red} < 2$ (see
Sect.~\ref{ss5:rm}). So, care must be exercised in interpreting the
polarization data from these pointing centers. On the other hand,
polarization angles show a linear variation over frequency in a large
part of the data, yielding good $RM$ determinations, which would not
be possible if large-scale offsets dominate. See Haverkorn et al.\
(2003a) for an extended discussion of this point. 

\section{Observational results}
\label{s5:res}

\subsection{Stokes $Q$ and $U$}
\label{ss5:qu}

\begin{figure*}
  \hbox{\psfig{figure=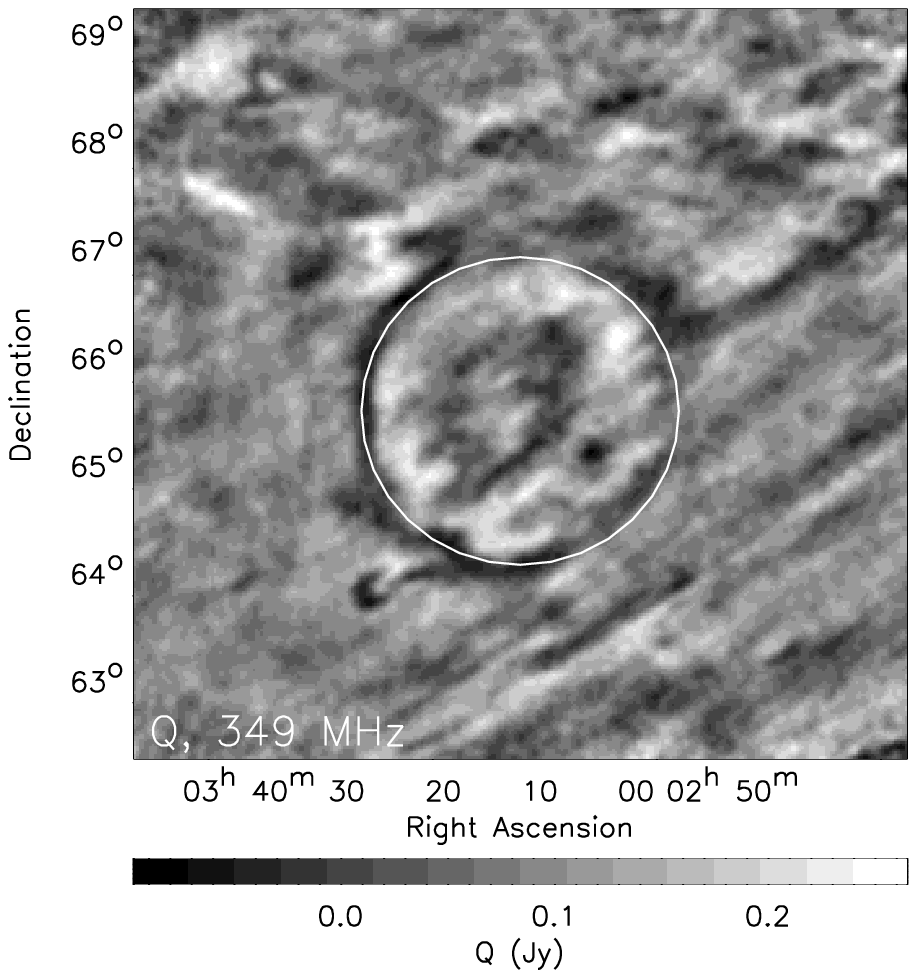,width=0.5\textwidth}
        \psfig{figure=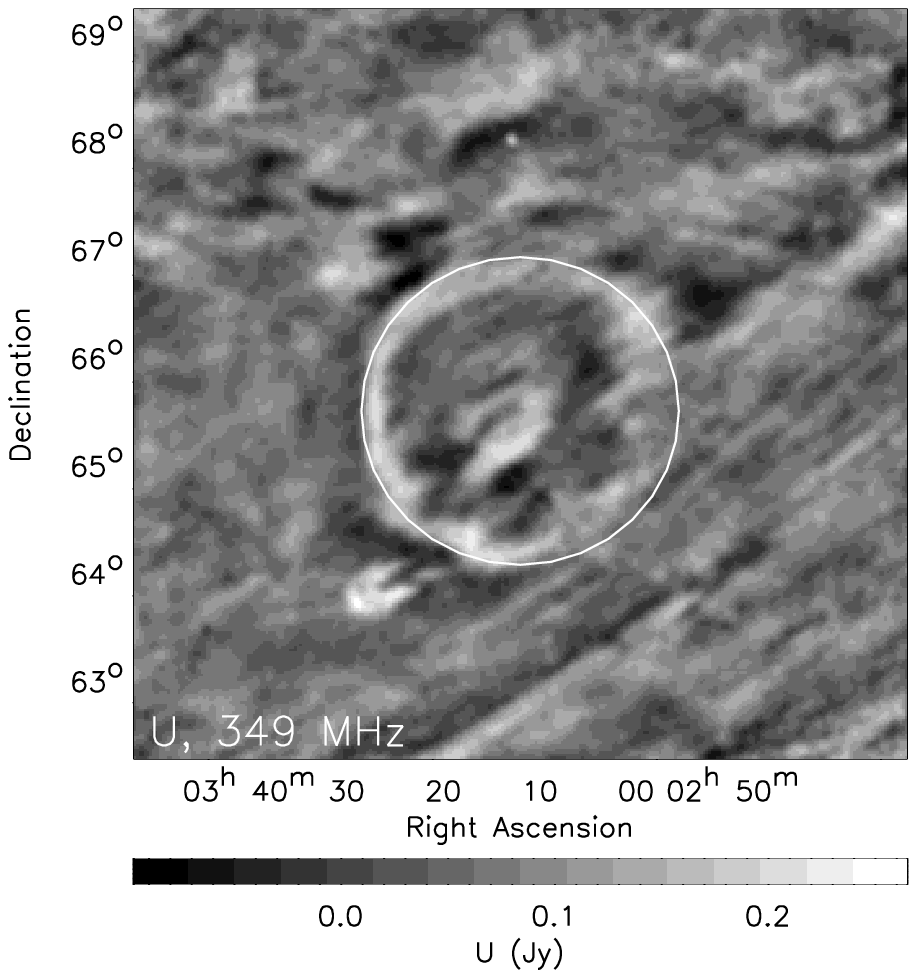,width=0.5\textwidth}}
  \caption{Stokes Q (left) and U (right) at 349 MHz, with a
           resolution of about 5\arcmin. A circle with radius 1.44\dg\
           and centered on $(\alpha,\delta) = (48.05\dg, 65.73\dg)$ is
           superimposed.} 
  \label{f5:qu}
\end{figure*}

The observed Stokes $Q$ and $U$ intensities at the 5\arcmin\
resolution are shown in Fig.~\ref{f5:qu}.  The distributions of $Q$
and $U$ values in the field are approximately Gaussian, centered
around zero, and have a width of 20 to 25~mJy/beam (equivalent to
polarized brightness temperatures $T_{b,pol}$ of 2.9~-~3.7~K) for the
five frequencies. The ring structure is clearly visible in $Q$ and
$U$. To emphasize the perfect circularity of the ring, a circle of
radius 1.44\dg\ and centered on $(\alpha,\delta) = (48.05\dg,
65.73\dg)$, which was fitted by eye to the ring-like structure in $Q$
and $U$, is superimposed in both maps. 

\subsection{Polarized intensity $P$}
\label{ss5:pol}

\begin{figure*}
  \centering
  \hbox{\psfig{figure=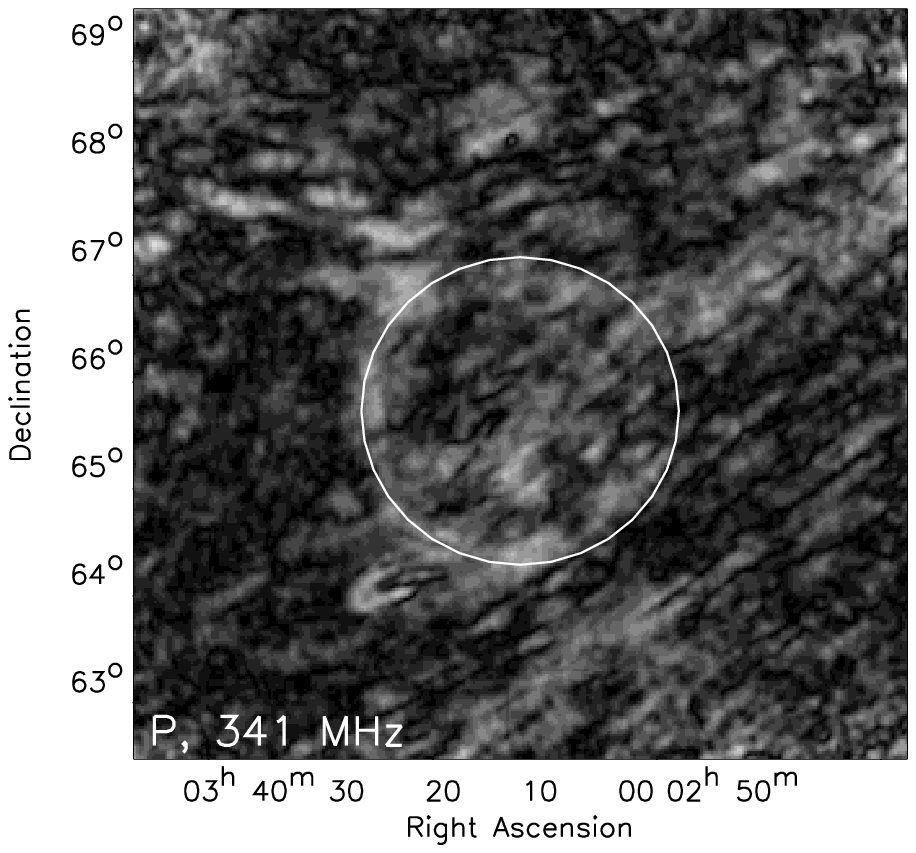,width=0.47\textwidth}
        \psfig{figure=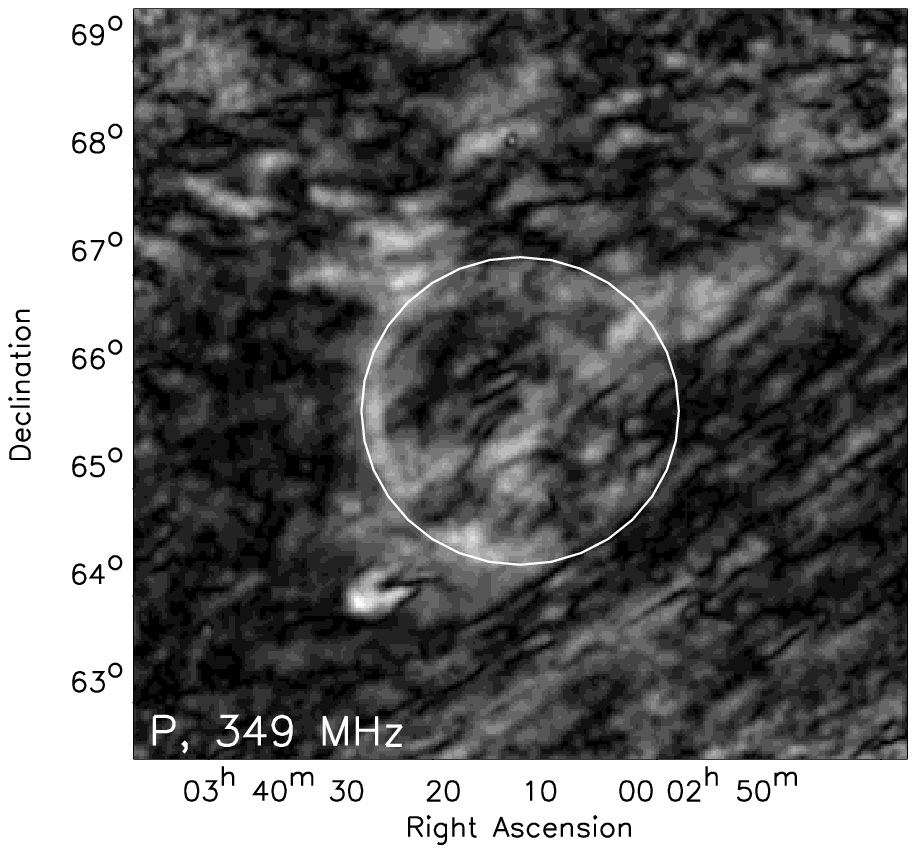,width=0.47\textwidth}}
  \hbox{\psfig{figure=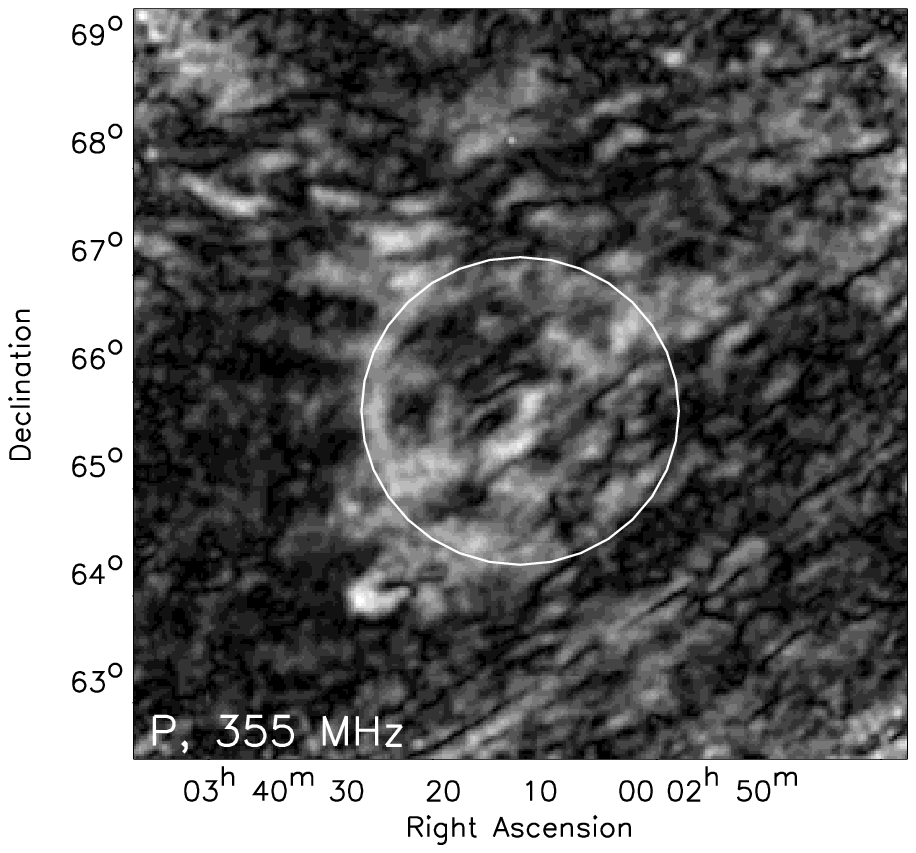,width=0.47\textwidth}
        \psfig{figure=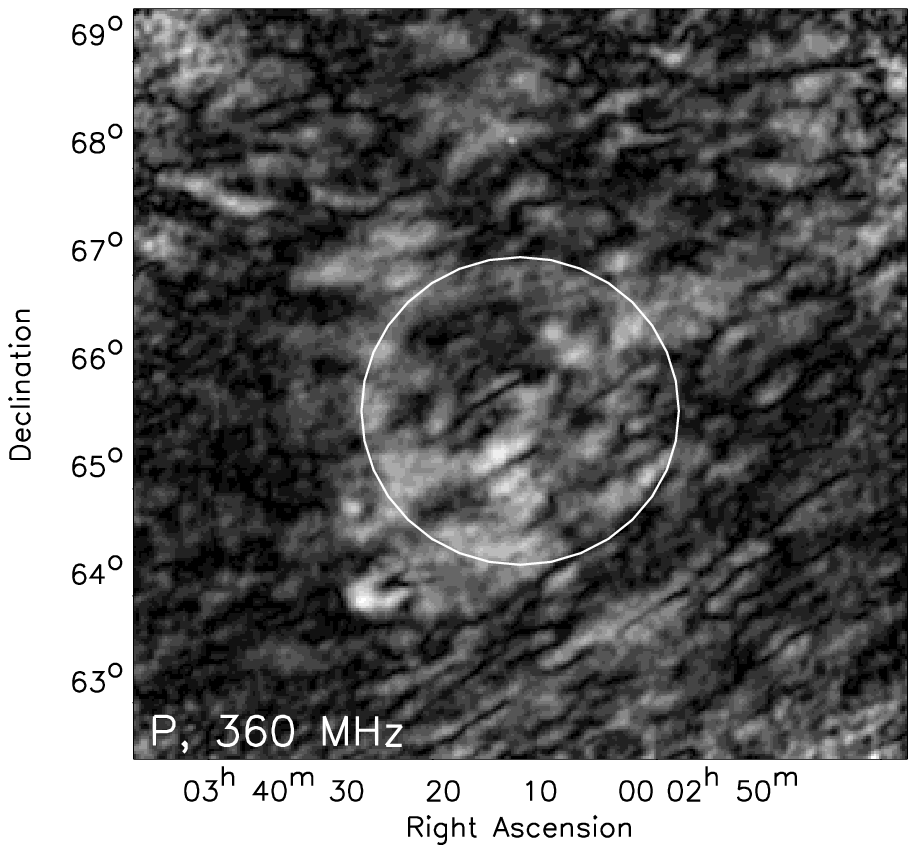,width=0.47\textwidth}}
  \hbox{\psfig{figure=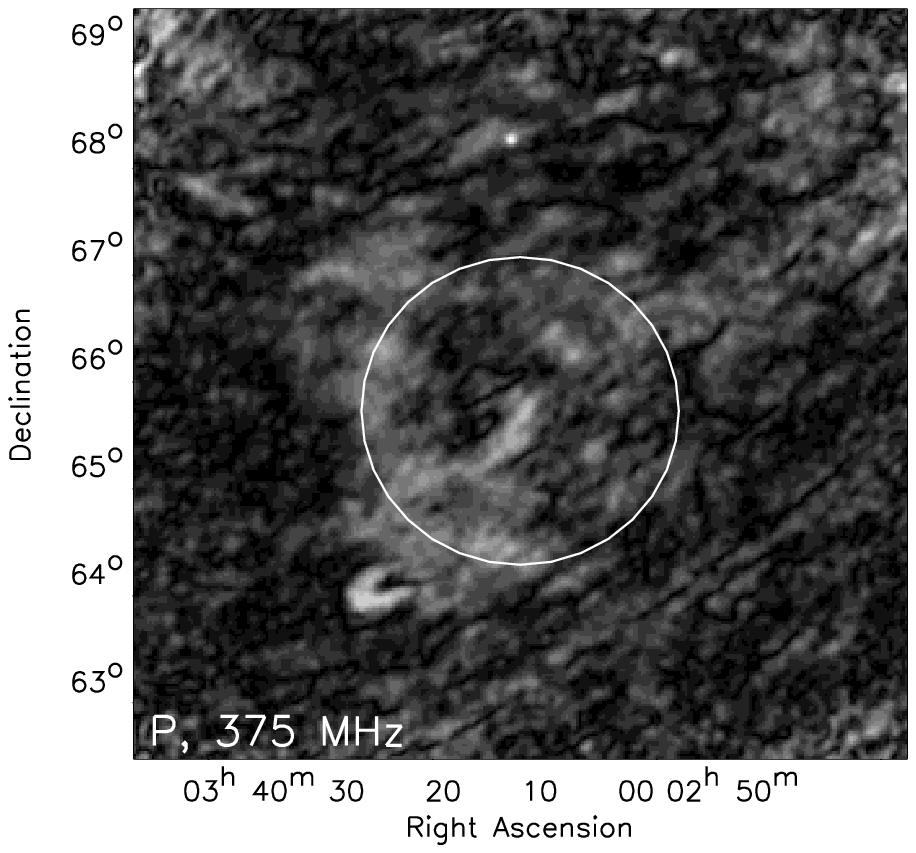,width=0.47\textwidth}
        \psfig{figure=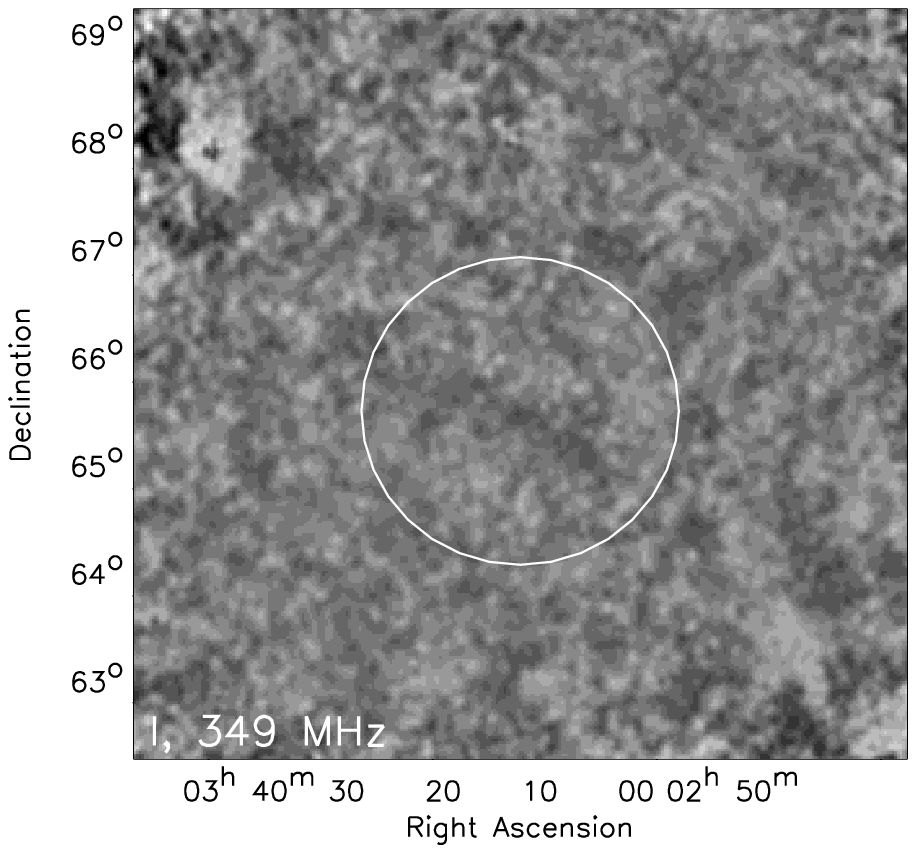,width=0.47\textwidth}}
  \caption{Polarized intensity $P$ for each frequency, at 5\arcmin\
         resolution, and total intensity~$I$ in the lower right plot.
         Superimposed are lines of constant Galactic latitude, and the
         same circle as in Fig.~\ref{f5:qu}.}
  \label{f5:pi}
\end{figure*}

The maps of polarized intensity $P$ derived from the Stokes $Q$ and
$U$ maps are shown in Fig.~\ref{f5:pi} for all 5 frequencies. White
denotes the highest intensity, and intensities above 110~mJy/beam
(equal to $T_{b,pol} = 16$~K) are saturated.  The maximum intensities
in the 5 frequency bands are 141, 122, 107, 120, and 143~mJy/beam,
respectively.  Superimposed are lines of constant Galactic latitude $b
= $3\dg, 5\dg, 7\dg, 9\dg, and 11\dg, and the superimposed circle is
the same as the one in Fig.~\ref{f5:qu}. Several regions of different
topology of polarized intensity are present in the field:
\begin{enumerate}
\item The most conspicuous feature is the ring in the center of the
      field. Although it is not as distinct as in polarization angle
      and Stokes $Q$ and $U$, the ring is still clearly visible in $P$
      over most of its circumference. Only in the southwest, the ring
      is not well defined. At 375~MHz, the ring seems to be more
      blurred than at the other frequencies.
\item A linear structure of high polarized intensity extends from
      the center of the ring towards the northwest, and is
      approximately aligned with Galactic latitude. We discuss this
      elongated structure of high $P$ in Sect.~\ref{ss5:bar}.
\item The southwestern corner of the field shows a filamentary pattern
      of ``canals'' of low~$P$ running from southeast to northwest,
      approximately along lines of constant Galactic latitude, that
      are remarkably straight over many degrees and always one
      beam width wide. This pattern is also visible in angle
      (Fig.~\ref{f5:pa}) and Stokes $Q$ and $U$ (Fig.~\ref{f5:qu}). 
      The southwestern part of the ring appears deformed in the
      direction of the filaments, but this deformation can be caused
      by averaging over the line of sight, so that the ring and
      filaments are not necessarily located at the same
      distance. Across the canals, there is an angle difference of
      90\dg\ (or 270\dg, 450\dg\ etc.).
\end{enumerate}

\subsection{Total intensity $I$}

In the lower right plot of Fig.~\ref{f5:pi} a map of total intensity
$I$ at 349 MHz is shown, from which point sources $> 5$~mJy/beam are
removed. The map has the same resolution of
5.0\arcmin$\times$5.5\arcmin\ and the same brightness scaling as the
$P$ maps in the other panels of the same figure.

No structure in total intensity $I$ is visible, although there is
abundant structure in polarization. The circular structure in the
upper left corner of the $I$~map is artificial and caused by a very
bright unpolarized extragalactic source at that position.  As an
interferometer acts as a high-pass filter, the map integral $I$
over the field is set to zero by lack of information about its true
level. From the single dish survey of Haslam et al.\ (1981, 1982) at
408 MHz, the total brightness temperature at 408 MHz at this position
is approximately 44.5~K with a temperature uncertainty of $\sim$~10\%
and including 2.7~K from the cosmic microwave background.  To obtain
the brightness temperature of the diffuse emission at 350~MHz, the
contribution of 2.7~K from the CMBR is subtracted, as well as the
contribution of 25\% from discrete sources, derived from source counts
(Bridle et al.\ 1972), assuming the spectral index of the Galactic
background and the extragalactic sources to be identical. The
intensities from the Haslam survey were scaled to our frequencies with
a temperature spectral index of $-$2.7. We thus estimate the total
brightness temperature at $\sim$~350~MHz to be 47~K.  The apparent degree
of polarization $p$ ($ = P/I$) is mostly above 100\%. This indicates
that $I$ is uniform on the scales that the WSRT is sensitive to, and
therefore the measured $I$ is close to zero. Small-scale structure in
polarization due to Faraday rotation and depolarization causes higher
$P$ than $I$, resulting in $p > 100$\%. 

\subsection{Polarization angle}
\label{ss5:pa}

\begin{figure}
  \centering
  \psfig{figure=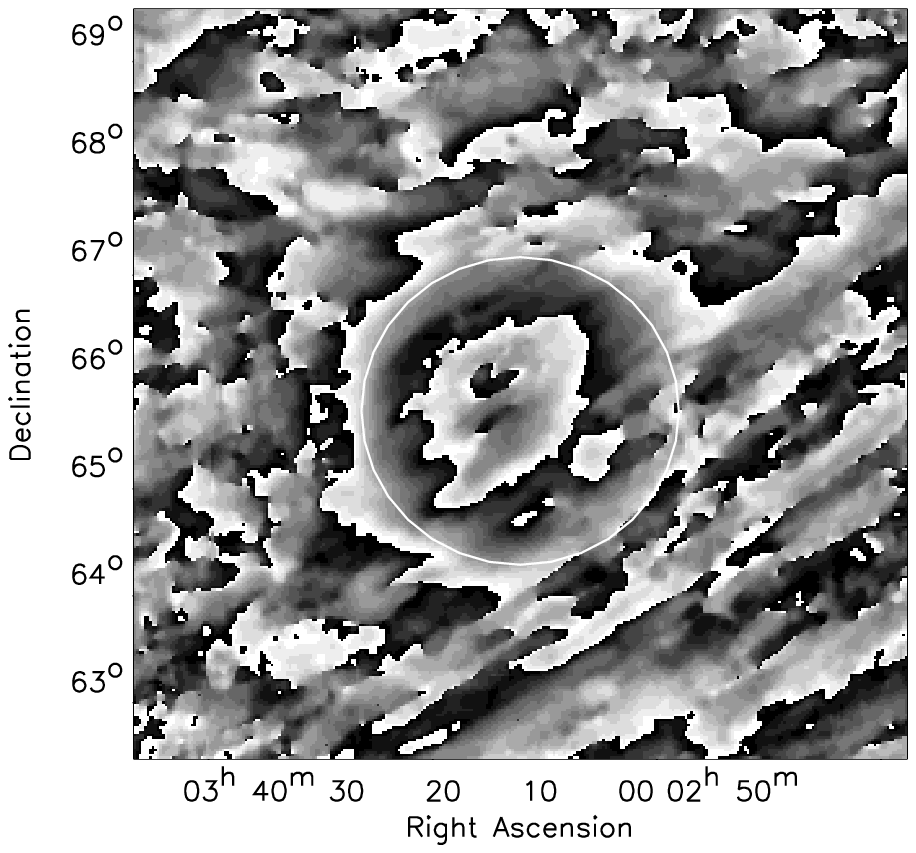,width=.5\textwidth}
  \caption{Polarization angle in the Horologium field at 349~MHz. The
      grey scale runs from $-90\dg$ to~90\dg, so that white is
      equivalent to black. The superimposed circle is the same as in
      Fig.~\ref{f5:qu}.}
  \label{f5:pa}
\end{figure}

In Fig.~\ref{f5:pa} we show a grey scale plot of polarization angle in
the Horologium region at 349~MHz. The range in angle is [--90\dg,
  90\dg], so that white denotes the same angle as black. The ring-like
structure is well visible, and is close to circular everywhere except
on the southwestern side.

\begin{figure}
  \centering
  \psfig{figure=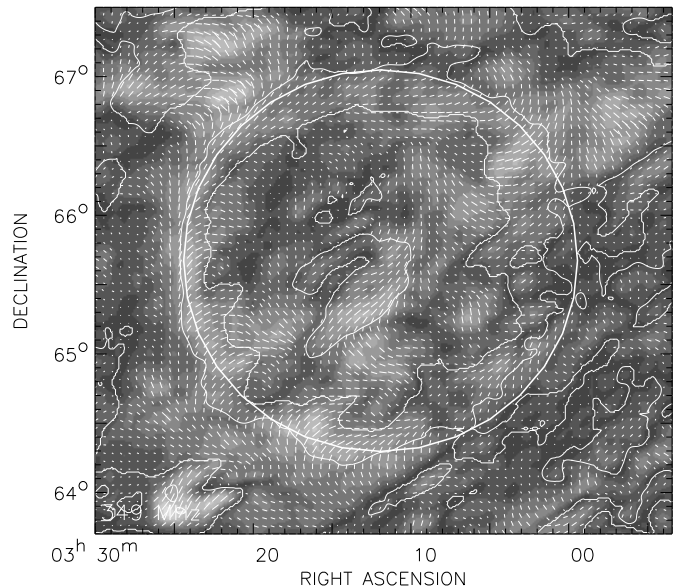,width=.5\textwidth}
  \caption{Detail of the Horologium field, where superimposed vectors
      are polarization vectors, and the grey scale is polarized
      intensity at 349~MHz. The superimposed circle is the same as in
      Fig.~\ref{f5:qu}. The contours delineate also polarization
      angle, showing that polarization angle is constant over the
      ring, except for the perturbation in the southwestern corner.}
  \label{f5:pa_ring}
\end{figure}

The variation of polarization angle across the ring is shown in
Fig.~\ref{f5:pa_ring}, where the grey scale is $P$ at a frequency of
349 MHz and the polarization (pseudo-)vectors are superimposed.  The
length of the vectors is proportional to $P$. The superimposed contours
are contours of polarization angle, and the circle is the same circle
as in Fig.~\ref{f5:qu}. The polarization angle is almost
perfectly constant over the surface of the ring, and only disturbed in
the southwest.

\subsection{Rotation measure}
\label{ss5:rm}

Rotation measures were calculated by linearly fitting the observed
polarization angle $\phi$ as a function of $\lambda^2$.  The polarization
vectors have an $n 180\dg$ ambiguity, but the derived rotation measures 
are so small that this ambiguity does not play a r\^ole in our
observations: $|RM| \la 10$ \radm, which means an angle change over
the full frequency range of 341~MHz to 375~MHz $\Delta\phi \la
75\dg$. Therefore, we calculate the $RM$ values  straightforwardly
with angle differences minimized.

\begin{figure}
  \centering
  \psfig{figure=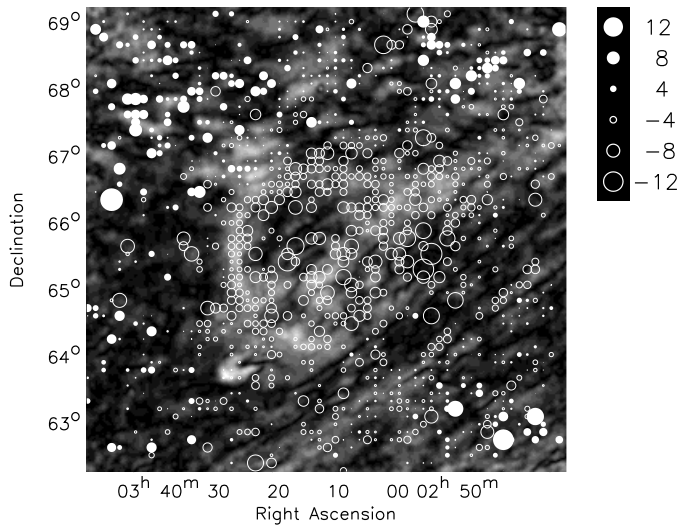,width=.5\textwidth}
  \caption{Rotation measure map. Circles represent $RM$s for which
           reduced $\chi^2 < 2$ and $\left<P\right> > 20$ mJy/beam
	   $\approx 4 \sigma$. The diameters of the circles scale
           with $RM$, and positive $RM$s are denoted by filled
           circles. Only one in four beams is shown.}
  \label{f5:rm}
\end{figure}

There are various mechanisms that can destroy the linear relation
between $\phi$ and $\lambda^2$ and make $RM$ determination unreliable,
such as depolarization (see Sect.~\ref{s5:or}) and the
insensitivity to large-scale structure (see Sect.~\ref{ss5:off}). We
have only taken into account those $RM$ values for which the linear
$\phi(\lambda^2)$-relation has reduced $\chi^2 < 2$ and where the
polarized intensity averaged over all wavelengths $\left<P\right> >
20$~mJy/beam ($\sim$ 4~$\sigma$). About 62\% of the polarized
intensity data has $\left<P\right> > 20$~mJy/beam, and approximately
29\% of these data (i.e. $\sim$~18\% of he total) also has a reduced
$\chi^2 < 2$. The resulting $RM$ map is given
in Fig.~\ref{f5:rm}. The circles denote valid $RM$ values according to
the above definition, where the diameter of the circle is proportional
to the magnitude of the $RM$. Filled circles denote positive rotation
measures. For clarity, only one out of two independent beams in both
directions is shown.

In the area enclosed by the ring all the $RM$s are negative, decreasing
towards the center of the ring to $RM \approx -8$ \radm. Further away
from the center of the ring, $RM$s increase up to a positive $RM$ of a
few \radm\ at the edges of the field.
These values are in agreement with earlier measurements at lower
resolution. In rotation measure maps produced by Bingham \&
Shakeshaft (1967) and by Spoelstra (1984), $|RM| \la 3$~\radm\ outside
the ring. They presented values of $RM$ inside the ring of $RM <
-5$~\radm\ and $RM < -3$~\radm, respectively.

\subsection{Extragalactic sources}

\begin{figure}
  \centering
  \psfig{figure=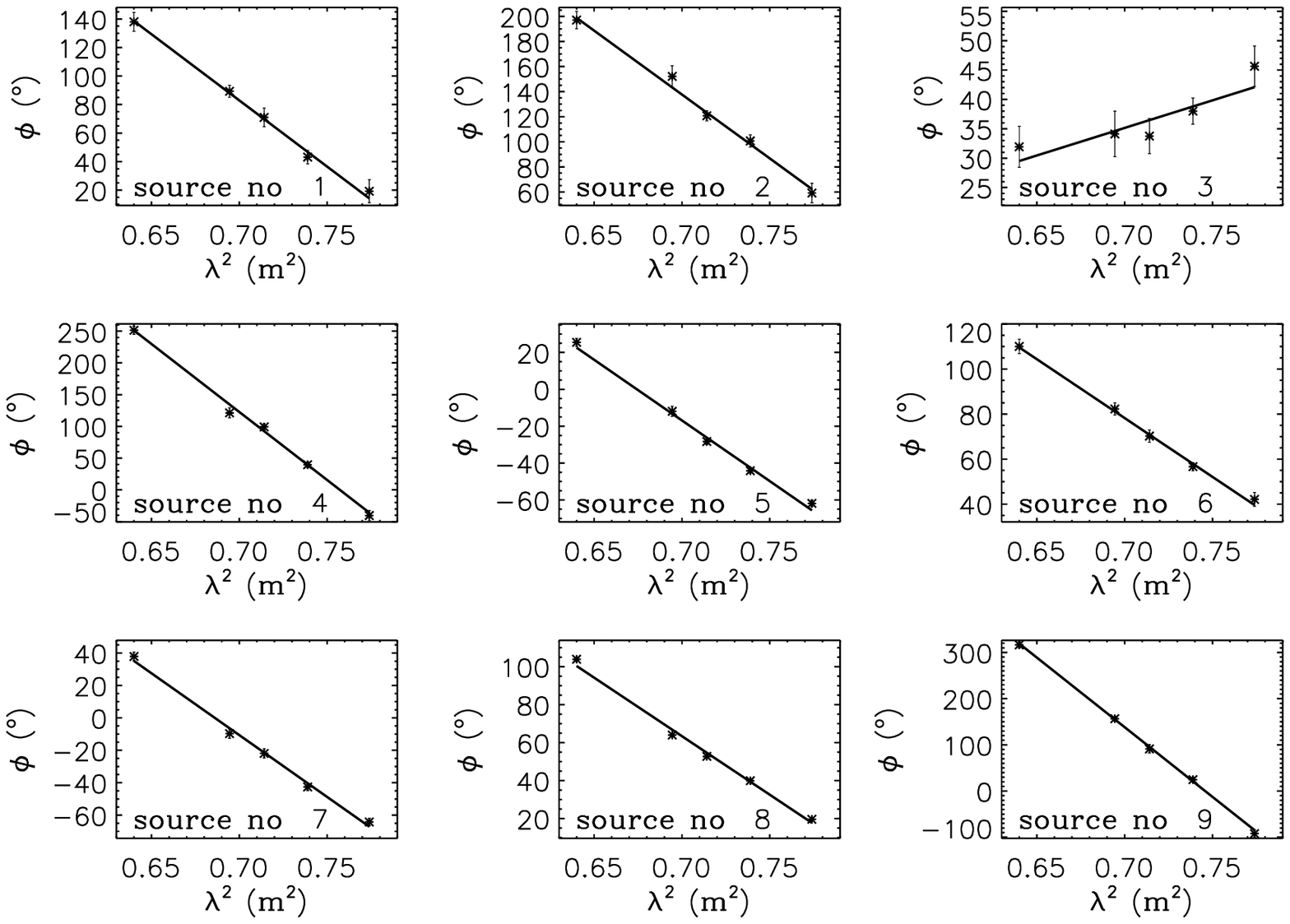,width=.5\textwidth}
  \psfig{figure=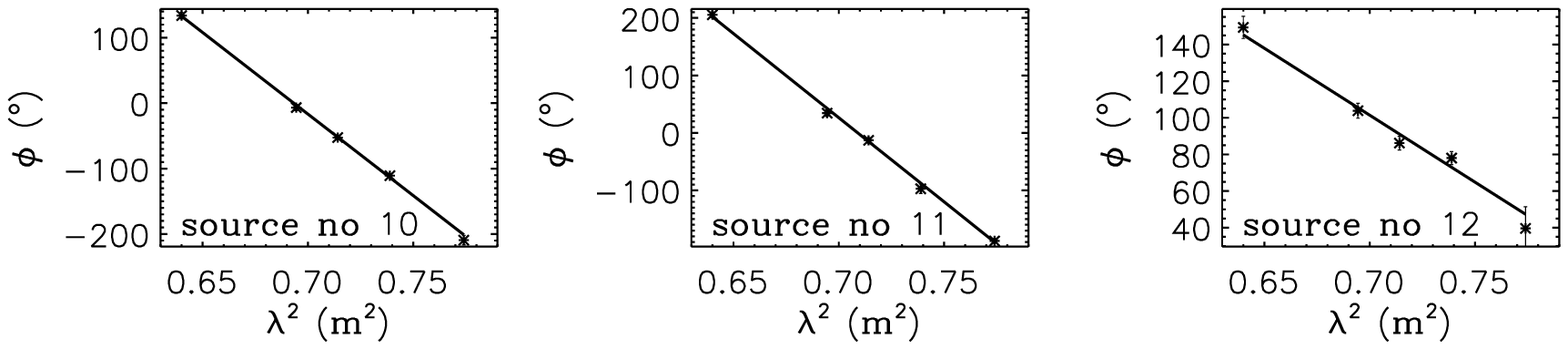,width=.5\textwidth}
  \caption{Graphs of polarization angle $\phi$ against $\lambda^2$
      for the 12 polarized extragalactic sources in the field. Note
      that the scaling on the $y$-axis is different for each plot.}
  \label{f5:src}
\end{figure}

\begin{figure}
  \centering
  \psfig{figure=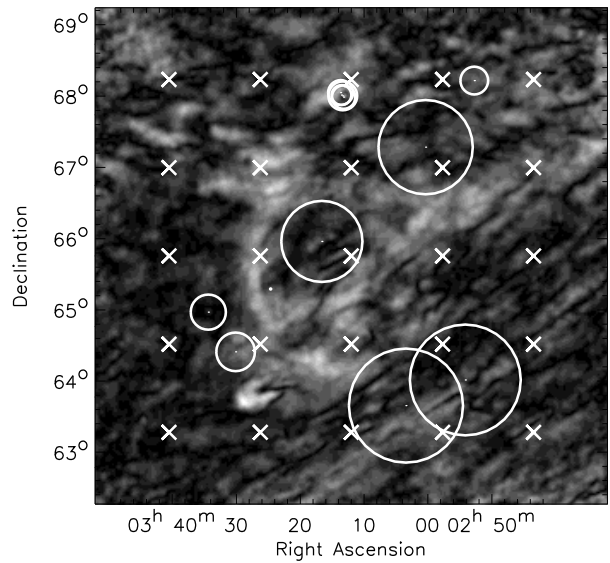,width=.45\textwidth}
  \caption{Rotation measures of polarized extragalactic sources
      indicated by white circles, overlaid on a grey scale
      representation of polarized intensity at 349 MHz. The $RM$
      scales with the diameter of the circles, where the extrema are
      $-52.6$ \radm\ and 1.6 \radm, and open circles denote negative
      $RM$. Instrumental polarization increases rapidly beyond the
      outermost pointing centers, denoted by crosses.}
  \label{f5:circle}
\end{figure}

Twelve polarized extragalactic sources were detected in the field. The
properties of the extragalactic sources were measured in maps with a
resolution of 1\arcmin. All selected sources have a polarized intensity greater
than 4~mJy/beam in each frequency band (the noise in the 1\arcmin\
resolution data is about 1~mJy/beam) and a degree of polarization
greater than 1\%. Near the outer edge of the outermost pointing centers,
instrumental polarization increases considerably, and sources in
that region were excluded. The characteristics of the polarized
extragalactic sources are shown in Table~\ref{table1}, the
$\phi(\lambda^2)$-fits in Fig.~\ref{f5:src}, and the positions of the
sources in Fig.~\ref{f5:circle}.

The brightest polarized source in the field with high polarization is
the giant double lobed radio galaxy WNB 0313+683 at $(\alpha,\delta) =
(3^h13^m, 68\dg20^m)$ (Schoenmakers et al.\ 1998).  In our analysis
this extended source was detected in four separate maxima,
viz. sources 5, 6, 7, and~8, all located at approximately the same
position in Fig.~\ref{f5:circle}. The high resolution measurements of
Schoenmakers et al.\ (15\arcmin\arcmin\ in the $RM$ map) show an
average $RM$ = $-$10.64~\radm, which they argue is Galactic, and a
residual $RM$ of about 2~\radm\ in each of the lobes. In our 1\arcmin\
resolution observations at 350~MHz, this would result in a variation
in polarization angle within the beam. Depolarization due to this
variation in polarization angle can destroy the linear relation
between $\phi$ and $\lambda^2$ and result in the relatively high
$\chi^2$ values in sources 5 and~8.

Source no.\ 3 is the only source with a positive $RM$, which is
derived from a good fit with $\chi_{red}^2 = 0.84$. As a check, taking
into account the $n\,180\dg$ ambiguity could not yield a $RM$ value
closer to that of the other sources. The next best fit has
$\chi_{red}^2 = 220$, so $RM$ = 1.6~\radm\ is indeed the only
acceptable $RM$ determination for this source.

\begin{table}
  \caption[]{Polarization data for extragalactic sources with
   measured polarization in the Horologium field. The second column
   gives position in $(h\!:\!m,\dg\!:\!m)$, and the third $RM$s with
   errors. Reduced $\chi^2$ of the $\phi(\lambda^2)$-relation is given
   in column 4. Columns 5, 6 and 7 give resp. $P$, $I$ (both in
   mJy/beam) and degree of polarization $p$ in percents, averaged over
   the five frequency bands.}
  \label{table1}
  \[
  \begin{array}{rcr@{\,\,\pm\,\,}lrrrr} \hline
    \noalign{\smallskip} \mbox{No.} & (\alpha,\delta) &
    \multicolumn{2}{c}{RM\; \mbox{(rad~m$^{-2}$)}} &
    \multicolumn{1}{c}{\;\;\;\chi^2} &
    \multicolumn{1}{c}{\;\;\left<P\right>\;} &
    \multicolumn{1}{c}{\;\;\left<I\right>\;} &
    \multicolumn{1}{c}{\;\;\left<p\right>} \\
    \noalign{\smallskip}
    \hline
    \noalign{\smallskip}
    1 \;\;& \mbox{[3:31,\,65:12]} & \;\;-16.2 & 1.2 
          &  0.37  &  4.8 & 168 &  \;\;2.9 \\
    2 \;\;& \mbox{[3:28,\,64:38]} & \;\;-17.8 & 1.2 
          &  0.67  &  4.5 & 253 &  \;\;1.8 \\
    3 \;\;& \mbox{[3:23,\,65:32]} & \;\;  1.6 & 0.6 
          &  0.84  &  8.7 &  97 & \;\; 9.0 \\
    4 \;\;& \mbox{[3:16,\,66:12]} & \;\;-37.4 & 1.2 
          &  1.67  &  4.4 &  73 & \;\; 6.0 \\
    5 \;\;& \mbox{[3:13,\,68:19]} & \;\;-11.5 & 0.3 
          &  3.55  & 15.5 &  84 & \;\;19.2 \\
    6 \;\;& \mbox{[3:13,\,68:17]} & \;\; -9.1 & 0.5
          &  0.47  & 10.4 &  88 & \;\;12.1 \\
    7 \;\;& \mbox{[3:13,\,68:16]} & \;\;-13.3 & 0.4
          &  1.61  & 10.6 &  72 & \;\;15.0 \\
    8 \;\;& \mbox{[3:13,\,68:15]} & \;\;-10.8 & 0.1
          & 13.34 & 37.3 & 366 & \;\;10.1 \\
    9 \;\;& \mbox{[3:05,\,63:53]} & \;\;-52.6 & 1.2 
          &  0.65  &  4.4 &  82 &  \;\;5.3 \\
    10\;\;& \mbox{[3:02,\,67:32]} & \;\;-43.5 & 0.9 
          &  1.05  &  6.4 & 218 &  \;\;2.9 \\
    11\;\;& \mbox{[2:57,\,64:15]} & \;\;-51.1 & 0.8 
          &  1.53  &  5.6 & 564 &  \;\;1.0 \\
    12\;\;& \mbox{[2:56,\,68:29]} & \;\;-12.8 & 1.1 
          &  1.56  &  5.5 & 238 &  \;\;2.3 \\
    \noalign{\smallskip}
    \hline
  \end{array}
  \]
\end{table}

\section{Observed properties of the ring-like structure}
\label{s5:prop}

The northeastern half of the ring is very regular in polarized
intensity, polarization angle and $RM$, whereas the southwestern part
appears to be influenced by the linear structure aligned with Galactic
latitude. Therefore, in analyzing the ring structure, we study azimuthal
averages of $P$, $\phi$ and $RM$ over position angles (N through E)
$-20\dg < \theta < 135$\dg, i.e.\ the undisturbed part of the ring.
The azimuthal averages are centered on $(\alpha,\delta) =
(3^h12.2^m,65\dg44^m)$, which is the center of the ring superimposed in
Fig.~\ref{f5:qu}.

\begin{figure}
  \centering
  \psfig{figure=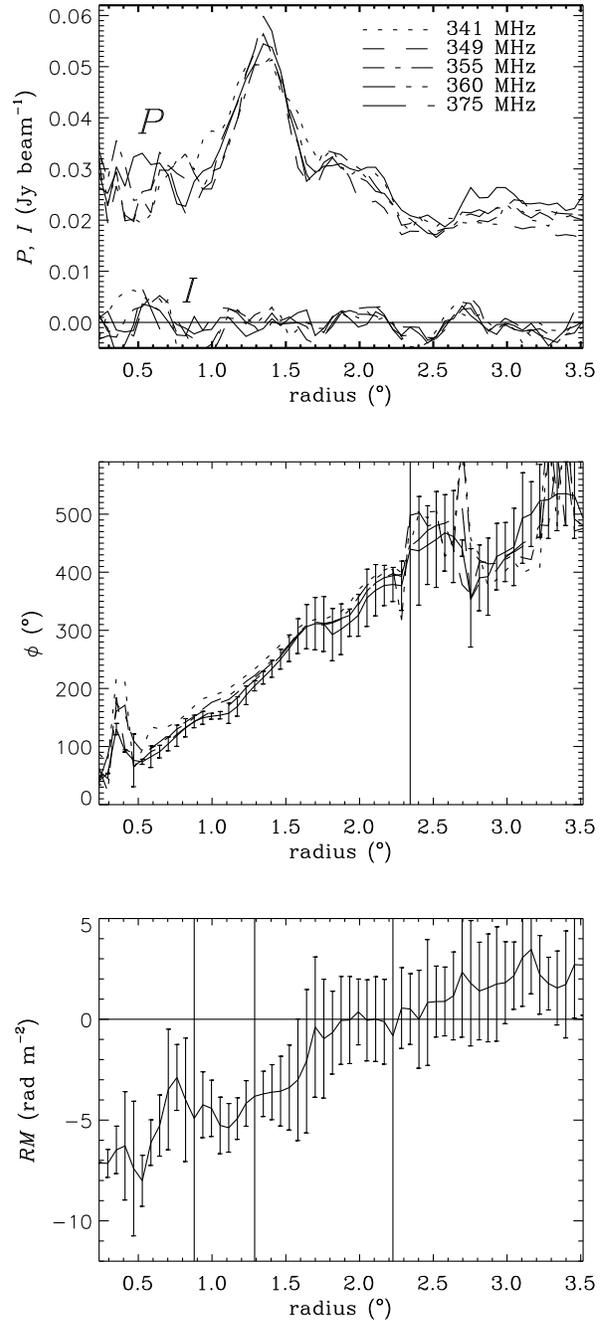,width=.45\textwidth}
  \caption{Azimuthal averages of polarized intensity $P$ and
      total intensity $I$ in Jy/beam for five frequencies (top),
      polarization angle $\phi$ in degrees for five frequencies
      (center) and $RM$ (bottom), averaged over the northern and
      eastern part of the ring structure (position angle $-20\dg <
      \theta < 135$\dg).} 
  \label{f5:ring}
\end{figure}

In Fig.~\ref{f5:ring}, we show the azimuthal averages of $I$,
$P$, $RM$ and $\phi$. The top panel shows $P$ and $I$ at 5
frequencies. The peak in $P$ at a radius of about 1.4\dg\ coincides
with the position of the ring. The polarized brightness temperature
$T_{b,pol}$ at that radius is approximately 8~K, $T_{b,pol}$ inside
the ring is about 3.8~K, and outside the ring about 3~K. The total
intensity $I$ is completely uncorrelated with $P$, and hardly exceeds
the noise. Note that the position and width of the peak in $P$ is
frequency-independent, while inside the ring the $P$ distribution does
vary with frequency.

The middle panel in Fig.~\ref{f5:ring} shows the weighted azimuthal
average of the polarization angle plotted as a function of radius,
again at 5 frequencies. The angles in each bin were chosen as close as
possible ($\pm$~n180\degr) to the mean angle value of the bin. Then,
the mean value was recalculated and the procedure was repeated until
the mean angle converged to a single value. The error bars indicate
standard deviations of the angle distribution, given for independent
beams only in the 341~MHz frequency band. The angle gradient is
remarkably linear between a radius of 0.6\dg\ and 1.7\dg, with some
structure at a radius of $\sim$~1\dg. Whereas $P$ has a maximum at
$\sim$~1.4\dg, the linear increase of polarization angle with radius
continues to at least 1.7\dg. This indicates that the structure or
feature responsible for the ring in $P$ may be actually larger than
the apparent size of the ring.

The bottom panel gives the azimuthally averaged observed $RM$ (only
values with $\chi_{red}^2 < 2$ and $\left<P\right> > 20$~mJy/beam) and
the standard deviations of the distribution of $RM$ over each
interval. The $RM$ does not vary much over the region at which the
peak in $P$ occurs. At radii larger than 1.7\dg, the $RM$ is slightly
positive: $0\la RM\la 2$~\radm. At smaller radii, $RM$ is negative,
decreasing to $RM\approx -8$~\radm\ in the center, with a small
increase around a radius $r = 0.7$\dg. This disk of negative $RM$ in
an environment of positive $RM$ is also observed in earlier studies
(Bingham \& Shakeshaft 1967, Spoelstra 1984). Although the general
shape of the $RM$ curve corresponds well to the shape of the
polarization angle curve, the shapes are different in detail, which we
attribute to an imperfect linear relation between $\phi$  and
$\lambda^2$.

The observations in $I$, $P$, $\phi$ and $RM$ impose stringent
constraints on the nature of the ring. The distribution of $RM$
provides the strongest constraint: $RM$ is positive outside the ring,
negative inside and the most negative at the center. The positive
$RM$ at large radii from the center of the ring is not necessarily
connected to the ring, but could also denote a background magnetic
field, of which the parallel component is directed towards the
observer. Although $RM$ structure is created by a combination of
structure in thermal electron density and magnetic field, a change in
sign of $RM$ always indicates a reversal of the parallel component of
the magnetic field.  So the magnetic field configuration that causes
the ring must reverse direction from outside to inside the ring, and
the magnetic field and/or electron density has to be the highest at
the center.  Furthermore, the lack of correlated $I$-structure implies
that the ring in $P$ cannot have been produced by emission. In the
next section, we first discuss what processes can cause the observed
distributions of $P$, $\phi$ and $RM$ in the ring. Subsequently, we
shall describe in Sect.~\ref{s5:conn} some known structures and
objects in the ISM, and discuss whether these are related to the ring
structure.

\section{The nature of the ring in $P$, $\phi$ and $RM$}
\label{s5:or}

From comparison of the $I$ and $P$ maps in Fig.~\ref{f5:ring}, it is
clear that the structure in $P$ cannot, even in part, be caused by
structure in $I$. Structure in $P$ can also be created by missing
large-scale structure in $Q$ and/or $U$, but in Sect.~\ref{ss5:off}
we have shown that in these observations missing large-scale structure
cannot dominate.  Therefore, the ring in $P$ is most likely due to a lack of
depolarization.  Several depolarization mechanisms can contribute to
create the ring in $P$. We shall discuss briefly the different
depolarization mechanisms thought to be of importance (for details see
Haverkorn et al.\ 2003a,b).

\subsection{Depth depolarization}

\begin{figure*}
  \centering
  \psfig{figure=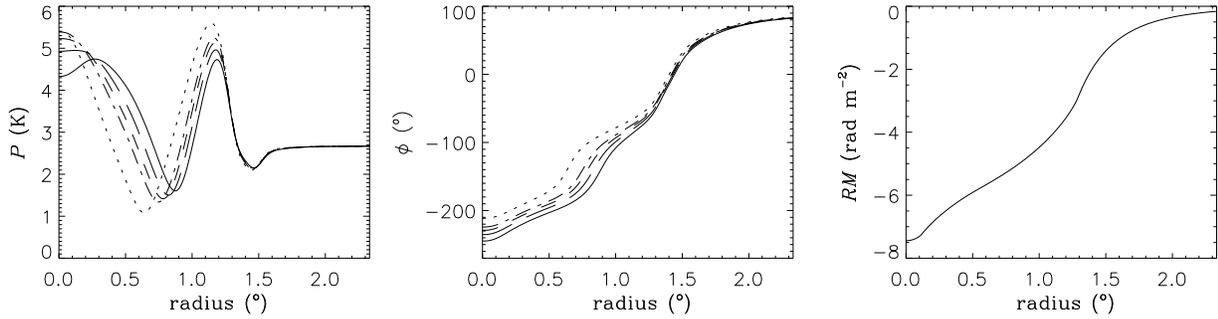,width=.9\textwidth}
  \caption{Azimuthally averaged $P$ (left), $\phi$ (center) and $RM$
      (right) plotted as a function of radius for a simple power law
      decrease of the $B$ and $n_e$ distributions. The different
      lines refer to the 5 frequencies.}
  \label{f5:sim}
\end{figure*}

Depth depolarization is defined as all depolarization processes
occurring along the line of sight and can be due to different physical
processes. First, if the magnetic field in a synchrotron emitting
medium has small-scale structure, then the emitted (intrinsic)
polarization angle of the synchrotron radiation will vary along the
line of sight, causing wavelength independent depolarization. 
Secondly, if the medium also contains thermal electrons, the
polarization angle of the radiation will be modulated 
by Faraday rotation as well, which causes additional depolarization
(internal Faraday dispersion). So small-scale structure in (parallel)
magnetic field and/or thermal electron density within the synchrotron
emitting medium causes small-scale depolarization.  These processes
were described analytically by Sokoloff et al.\ (1998) for several
different geometries of the medium, and numerically in Haverkorn et
al.\ (2003b) using observational constraints.

We modeled the effect of depth depolarization in the observations of
the ring-structure using simple distributions of electron density
$n_e$ and magnetic field $B$ on a rectangular grid. These
distributions are not self-consistent, but the only goal of this
simple model is to obtain a $P$ and $\phi$ distribution that is
similar to the observations, i.e.\ approximately linear in $\phi$ and
ring-like in $P$. Synchrotron radiation of emissivity $\varepsilon
\propto B_{\perp}^2$ is emitted in the regions where $B_{\perp}$ is
non-zero, and is Faraday-rotated while propagating through the medium,
depending on the local $B_{\pl}$ and $n_e$ distributions.
Furthermore, a polarized background contribution $P_b$ is added, which
is also Faraday-rotated.  Both magnetic field components, parallel and
perpendicular to the line of sight, and the electron density
distribution were assumed to decrease as a power law outwards. The
specific values of the power law indices were chosen so that the
observed $\phi$ and $RM$ distributions were approximately reproduced
(Fig.~\ref{f5:sim}), i.e.\ the magnetic field decreases as $r^{-5}$
(only at $r > r_0$, where $r_0$ is a free parameter too)
and the electron density decreases as $r^{-0.4}$.

This figure shows the model output $P$ and $\phi$ at 5 frequencies,
and $RM$. We have chosen $B_{\pl} = -3.5~\mu$G, $B_{\perp} = -2~\mu$G,
$n_e = 0.2$~cm$^{-3}$, and $P_b = 4$~K. This reproduces the shape and
magnitude of $\phi$ and $RM$ reasonably well, but the $P$ distribution
is very different from the observed one. First, the predicted $P$ at
the center is much larger than is observed. However, this discrepancy
could be explained by assuming a chaotic magnetic field component at
the center of the circle (without worrying yet what this could mean
physically). This would result in depolarization and a lower observed
$P$ in the center of the modeled circle.

However, a more severe problem is posed by the wavelength dependence
of the model predictions. Although our models have very different $B$
and $n_e$ distributions and either spherical or cylindrical symmetry,
they all show a distinct wavelength dependence of the peak in $P$, as
in Fig.~\ref{f5:sim}. But from the observations, the position of the
peak in $P$ does not change with wavelength, see Fig.~\ref{f5:ring}.
The wavelength dependence of $P$ appears to be a generic property of
all models involving depolarization due to depth depolarization.
However, the mechanism that can create wavelength independent
depolarization, viz.\ tangled magnetic fields, yields structure in
$I$, contrary to what is observed. Therefore we conclude that depth
depolarization cannot be the main process that creates the ring in
$P$, although we do expect depth depolarization to be present, e.g.\
in depolarizing the background.

\subsection{Beam depolarization}

Beam depolarization, i.e.\ the averaging out of polarization vectors
within one synthesized beam, is significant in the field. As there is
structure in $RM$ on beam scales, it is likely that $RM$ varies on
scales smaller than the beam as well. Furthermore, at the positions of
the depolarization canals, the influence of beam depolarization is
clearly visible, see Sect.~\ref{ss5:can}. (Partial) beam
depolarization can destroy the linear $\phi(\lambda^2)$-relation, but
does not necessarily do so. At low polarized intensities, the
influence of beam depolarization can be considerable, and observed
$RM$ values at low polarized intensity should be used with care, as
they can deviate from the true $RM$ value.

Beam depolarization, due to chaotic structure in polarization angle on
scales smaller than the beam, can arise due to tangled magnetic fields
and/or small-scale variations in thermal electron density. A possible
explanation for the lack of $P$ in the central part of the ring could
be a chaotic magnetic field in the center, while the outer parts of
the ring must exhibit very coherent magnetic fields and electron
density. However, although $P$ drops towards the center of the ring,
polarization angles are coherent over the whole structure, indicating
that beam depolarization due to chaotic sub-beam-scale structure does
not dominate.

\begin{figure}
  \centering
  \psfig{figure=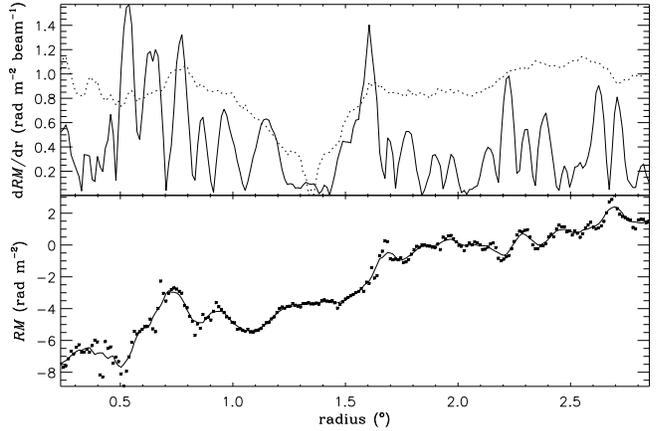,width=.5\textwidth}
  \caption{Top plot: derivative of azimuthally averaged $RM$ with
           respect to radius $r$. The solid line denotes the observed
           $dRM/dr$, and the dotted line is $dRM/dr$ derived from the
           observed polarized intensity, using Eq.~(\ref{e:graddep}). 
	   Bottom plot: observed azimuthally averaged $RM$ in
           dots. The superimposed line is the $RM$ smoothed over about
           1.5 beams which is used for the estimate of $dRM/dr$.} 
  \label{f5:graddep}
\end{figure}

Beam depolarization can also be due to a spatial gradient in $RM$. For
 a gradient $dRM/dr$ over many beams, the depolarization factor $p$ is
 (Gaensler et al.\ 2001)
\begin{equation}
   p = \frac{P_{\scriptstyle\rm observed}}{P_{\scriptstyle\rm original}} 
     = \,\exp\left[ -\frac{1}{\ln 2}\left(\frac{dRM}{dr}\right)^2
                     \lambda^4 \right]
   \label{e:graddep}
\end{equation}
where $dRM/dr$ is the gradient of $RM$ over the beam and $r$ is the
radius of the ring, expressed in the number of beams. In the upper
panel of Fig.~\ref{f5:graddep}, we compare the observed radial
gradient in $RM$ (solid line) with the theoretical $RM$ gradient from
Eq.~(\ref{e:graddep}) (dotted line), using the observed polarized
intensity $P$ at 341~MHz. Using $P$ from other frequency bands gives
very similar results. The derivative of the observed $RM$ was computed
after smoothing $RM$ by about 1.5 beams (lower panel). At the position
of the ring, the modeled $RM$ gradient shows the same decrease as the
observed gradient. Furthermore, at the position where the gradient in $RM$
increases, the depolarization also increases. This indicates that the
high $P$ within the ring, as well as the decrease in $P$ at the inner
and outer boundaries, can be caused by the gradient in $RM$.  However,
in the center of the ring, as well as outside the ring, the existence
of a gradient in $RM$ would result in a lower depolarization than
is observed. So both at the center and outside the ring, other
depolarization mechanisms contribute significantly.

\subsection{The nature of the ring}

The distance to the ring-like structure is not constrained.  Depth and
beam depolarization introduce a Faraday depth or polarization
horizon, because polarized radiation emitted at large distances is
more likely to be depolarized than radiation emitted nearby. But because the
foreground and background of the ring must be very uniform, the
polarization horizon can be at very large distances. However, if the
ring is located at large distances, say in or behind the Perseus arm,
it would require a path length of more than 2~kpc with an unusually
uniform magnetic field and electron density along the path
length. Moreover, if the ring is in the Perseus arm, its size would be
$> 100$~pc, which is unlikely in view of the regular shape of the
ring. Therefore, the ring is unlikely to be located in (or behind) the
Perseus arm and is probably an inter-arm feature.

The reversal in the direction of the magnetic field from outside to
inside the ring, and the $RM$ that becomes more negative towards the
center, provides strong constraints for possible explanations for the
ring, because they would not allow magnetic field configurations that
have spherical symmetry (such as stellar winds or supernovae). This is
because in those configurations, the $RM$ contribution from the front
half of the structure compensates the (opposite) $RM$ contribution
from the far end (e.g.\ radial fields in young supernovae, bubbles
blown in a regular magnetic field perpendicular to the line of
sight). Alternatively, the $RM$ would be higher at the edges of the
ring than in the center (e.g.\ bubbles blown in a regular magnetic
field parallel to the line of sight). Instead, the structure that
creates the ring-like structure, must have a magnetic field directed
away from us, in surroundings where the magnetic field has a small
component towards us. Furthermore, the magnetic field strength and/or
electron density must increase from the edge towards the center of the
ring. These constraints make a connection of the ring with many known
classes of objects unlikely.  Below we discuss the plausibility that
some known gaseous structures in the ISM could be responsible for the
polarized ring.

\section{Connection of the ring with known ISM objects}
\label{s5:conn}

In this section, we discuss several known objects in the ISM to which
the ring could possibly be connected. We conclude that the observed
$RM$ structure makes a connection to any known ISM object unlikely.

\paragraph{Planetary nebula}
The circular form of the ring structure suggests it might be due to a
planetary nebula (PN). The radio emission from a PN is free-free
emission, which is negligible at our low frequencies for any
reasonable temperature.  However, the ring has an angular diameter
more than ten times as big as any other PN observed before. If the PN
were 5~pc in size, about the size of the largest ancient PN known
(Tweedy \& Kwitter 1996), then it would be at a distance of about 100
pc and it is unlikely that it hasn't been observed before.
Furthermore, it is difficult to see how the magnetic field
configuration needed to create the ring could be present in a PN.  For
these reasons we believe it unlikely that the ring structure is a
planetary nebula.

\paragraph{Str\"omgren sphere from HD20336}
Close to the projected center of the ring the B2V star HD20336 is
located at a distance of 246~$\pm$~37 pc (Hipparcos Catalogue,
Perryman et al.\ 1997). Verschuur (1968) suggested that the ring could
be related to this star. If the ring-structure in $P$ coincides with
the Str\"omgren sphere of HD20336, its radius would be
$R_s$~=~6~pc. The excitation parameter $U = R_s \, n_e^{2/3}$ is
2.6~pc~cm$^{-2}$ for a B2V star (Panagia 1973), which yields an
electron density inside the Str\"omgren sphere of $n_e \approx
0.1$~cm$^{-3}$. The emission measure $EM = \int n_e^2\,dl = 2\,
U^3/R_s^2$ solely from the Str\"omgren sphere is 0.96~cm$^{-6}$~pc
($\sim$~2~R), which could have been detected in the Northern
Sky Survey  of the Wisconsin H$\alpha$ Mapper (WHAM, Haffner et al.,
in prep., Reynolds et al.\ 1998), but wasn't (see
Fig.~\ref{f5:wham}). In addition, the Str\"omgren
sphere explanation is unlikely because of the high proper motion of
the star, viz.\ 18~mas/yr, which is equivalent to 5\dg\ in a Myr. The
recombination time $t_r \approx (\alpha^{(2)}\, n_e)^{-1}$, where the
recombination coefficient to the second level of the hydrogen atom
$\alpha^{(2)} \approx 1.4 \times 10^{-13}$. So the recombination time
is 8 Myr with $n_e = 0.1$~cm$^{-3}$. Due to the high proper motion of
the star, we conclude that a circular Str\"omgren sphere cannot be
maintained. Rather, the structure would be elongated in the opposite
direction to the proper motion of the star. In addition, it is
difficult to imagine a Str\"omgren sphere with a sufficiently
asymmetric magnetic field structure to explain the observed $RM$, as
described above.  Hence, the ring is unlikely to be a Str\"omgren
sphere around the star HD20336.

\paragraph{Stellar wind from HD20336}
We use the interstellar wind-blown bubble model by Weaver et al.\
(1977) to estimate the time needed to blow a bubble of radius 6~pc
with a stellar wind from a B2V star. Weaver et al.\ estimate the
radius of a blown bubble as
\[
R_s = 26.5 \;(L_{36}\, n_H^{-1}\, t_6^{\,3})^{0.2} \mbox{ pc}
\]
where $L_{36}$ is the mechanical energy of the stellar wind in units
of 10$^{36}$~erg~s$^{-1}$, $n_H$ is the original hydrogen particle
density before passage of the shell, and $t_6$ is the elapsed time in
Myrs. Assuming that $L_{mech} = 2.5\times 10^{-3} L_{bol}$ (Israel \&
van Driel 1990), and that the absolute luminosity of a B2V star is $L
= 3.7 L_{\odot}$ (Panagia 1973), then $L_{36} =
2.2\times10^{-4}$~erg~s$^{-1}$. With $R_s = 6$~pc and $n_H =
0.1$~cm$^{-3}$, the time needed to blow a bubble of the size of the
ring is almost 5~Myr. If $n_H$ is ten times lower, $t_6$ would still
be 2.2~Myr. Compared to the proper motion of the star, the time needed
to blow a circular bubble is far too long to produce the observed
structure. Therefore, the ring cannot be due to the effect of a
stellar wind either.

\paragraph{Supernova remnant}
For supernova remnants, a $\Sigma$-D relation exists between
brightness and distance. Using the non-detection of signal in $I$ as
an upper limit for the emission of the ring structure, the $\Sigma$-D
relation of Case \& Bhattacharya (1998) would imply a distance larger
than 36~kpc if the ring was a SNR. Clearly the ring structure is much
too faint in $I$\/ to be related  to a supernova remnant.

\paragraph{Superbubble or chimney}
Superbubbles that become chimneys as they blow out  material from
the spiral arm where they originate, have been observed in \HI, not only
into the Galactic halo, but also into the inter-arm region in the
Galactic plane (McClure-Griffiths et al.\ 2002). The ring could be
such a chimney blown away straight from the Sun. Galactic magnetic
field frozen in in the plasma that is blown out can cause the observed
high negative $RM$. However, if this is the case, one would expect the
magnetic field to be maximal at the edges of the chimney, and a low
electron density inside the chimney.

\section{The ring as a magnetic structure}
\label{s5:magn}

It is possible that the ring is created through bending of magnetic
field lines. E.g. contraction and motion of a plasma cloud or an
analogue to the Parker instability can enhance and reverse magnetic
fields.  The $RM$ plot in Fig.~\ref{f5:ring} shows an increase in $RM$
out to a radius of about 3\dg. If the azimuthal average over $RM$ is
taken over 360\dg, $RM$ still increases slowly to 0~\radm\ at $r =
3$\dg\ and to 2~\radm\ at $r = 4$\dg. This behavior is more like a
magnetic structure, with decreasing magnetic field over a large
radius, than like a structure in thermal electron density which has
more or less definite boundaries.  We conclude that the ring-like
structure in $P$ is most likely a funnel-shaped enhancement of magnetic
field, directed straight away from us, possibly related to magnetic
flux tubes (Parker 1992, Hanasz \& Lesch 1993). 
A ``magnetic anomaly'' of a few degrees in size has been found in the
direction $\ell \approx$~92\dg, $b \approx$~0\dg\ (Clegg et al. 1992,
Brown \& Taylor 2001) from $RM$s of extragalactic sources. However,
$RM$s of extragalactic sources in the direction of the ring do not
show a local magnetic field reversal (Fig.~\ref{f5:circle}). Therefore
we expect the ring-structure to be more localized or less strong than
this magnetic anomaly. The magnetic ring-structure may very well
be accompanied by an increase in electron density.
 
Although the distance and shape of the ring-structure are very
uncertain, we can try to give estimates of the necessary magnetic
field and electron density changes. An upper limit for the change in
$n_e$ is given by the non-detection of the ring in the Northern
Sky Survey  of the Wisconsin H$\alpha$ Mapper (WHAM). The WHAM can detect an
H$\alpha$ intensity of 0.05~R (1~R~=~1~Rayleigh is equal to a
brightness of $10^6/4\pi$ photons cm$^{-2}$ s$^{-1}$ ster$^{-1}$, and
corresponds to an emission measure $EM$ of about 2 cm$^{-6}$ pc for
gas with a temperature T = 10000~K). So
\begin{equation}
  \Delta (n_e^2 \, dl) = 2 \,n_e \, \Delta n_e \, dl < 0.1 \;\;
                         \Rightarrow \;\;  \Delta n_e\, dl < 0.5
  \label{eq5:ne}
\end{equation}
assuming a constant path length over the field and a background $n_e =
0.1$~cm$^{-3}$. The change in $RM$ is 
\begin{eqnarray}
  \Delta RM &=& 0.81 \,\Delta (n_e B_{\pl} dl) \nonumber    \\     
            &=& 0.81 (\,n_e \, \Delta B_{\pl} \, dl 
                    + \Delta n_e\, B_{\pl}\, dl +
                      \Delta n_e \,\Delta B_{\pl}\, dl)
  \label{eq5:rm}
\end{eqnarray}
The change in $RM$ from inside to outside the ring is $\Delta RM
\approx -8$~\radm, and assuming that the background magnetic field
$B_{\pl} \le 1 \mu$G, combination of Eqs.~(\ref{eq5:ne})
and~(\ref{eq5:rm}) yields
\begin{equation}
  \Delta B_{\pl}\, (dl + 5) \ga 100 
\end{equation}
If $D$ is the distance to the ring, and $f$ is the axial ratio of
the object with its major axis along the line of sight, then $dl =
D f \tan(3\degr)$. Estimates of $n_e$ and minimum $B_{\pl}$ for
different $D$ and $f$ are given in Table~\ref{t5:estbne}.

\begin{table}
  \centering
  \begin{tabular}{c|c|c|c|c}
    $D$ (pc) & $f$ & $\Delta B_{\pl}$ ($\mu$G) &
    $\Delta n_e$ (cm$^{-3}$) &  $dl$ \\
    \hline\hline
    100  &	1 & $\ga$ 10  &	0.1   &	5   \\
    500  &	1 & $\ga$ 3   &	0.02  & 25  \\
    1000 &	1 & $\ga$ 2   &	0.01  &	50  \\
    100  &	5 & $\ga$ 3   &	0.02  &	25  \\
    500  &	5 & $\ga$ 0.8 &	0.004 &	125 \\
    1000 &	5 & $\ga$ 0.4 &	0.002 &	250 \\
  \end{tabular}
  \caption{Estimated values of $\Delta B_{\pl}$ and $\Delta n_e$
    between ring structure and background, for different
    estimates of distance $D$ and axial ratio $f$.}
  \label{t5:estbne}
\end{table}

If the ring is located at very large distances (say $\ga 500$~pc), its
width becomes so large that the regularity of the ring becomes hard to
explain. A nearby spherical structure ($f=1$) needs a large change in
magnetic field, according to Table~\ref{t5:estbne}, whereas an
elongated structure with a smaller magnetic field change can also
accommodate for the observed $RM$ change. In any case, the change in
electron density in the ring is small, due to the upper limit imposed
by the non-detection of the ring in H$\alpha$.

\subsection{Are there more ring-like structures?}

\begin{table*}
  \centering
  \begin{tabular}{l|c|c|c}
    & Gray et al.\ 1998 & Uyan\i ker \& Landecker 2002 & this paper \\
    \hline\hline
    $(l,b)$          & (137.5\dg, 1\dg)       & (91.8\dg, --2.5\dg)
                     & (137\dg,7\dg) \\
    size             & 1\dg$\times$2\dg       & (1.5 - 2)\dg$\times$2\dg
                     & 2.8\dg$\times$2.8\dg \\
    $\Delta\phi$     & 280\dg                 & 100\dg  & 300\dg \\
    $\Delta RM$      & 110 \radm              & 40 \radm & 10 \radm \\
    $\phi$ structure & linearly increasing    & ``linearly'' decreasing
                     & linearly decreasing \\
    $P$ structure\hspace*{1cm} & ring         & ring & ring \\
    distance         & 440~pc $< d <$ 1.5 kpc & 350 $\pm$ 50 pc & ? \\
    \multicolumn{4}{c}{\mbox{}}\\
  \end{tabular}
  \caption{Comparison of the ring discussed here with two earlier
  detections of circular or elliptical polarization structures by Gray
  et al.\ (1998) and Uyan\i ker \& Landecker (2002).} 
\end{table*}

There have been two earlier detections of elliptical structures in
polarization angle and polarized intensity without correlated
structure in total intensity published, both in single-frequency
observations at 1420~MHz. The first detection by Gray et al.\ (1998)
is an elliptical feature of about 2\dg$\times$1\dg\ in size, which
shows a linear increase in polarization angle towards the
center. Polarized intensity shows a ring-like behavior, and the
regular structure in $\phi$ extends to a larger radius than where $P$
peaks.

The ``Polarization Lens'' observed by Uyan\i ker \& Landecker (2002)
shows approximately similar characteristics: an approximately linear
decrease of polarization angle towards the center, and a ring-like
structure in $P$, although the regular structure in angle seems to
trace an ellipse rather than a circle.  The characteristics in $P$ and
$\phi$ are so similar for the three features, that it is tempting to
consider them as members of the same class of objects.

However, although the appearance of the three rings is similar, the
changes in $RM$ responsible for the angle changes are vastly
different. Although the first two objects are only detected at one
frequency, and therefore their $RM$ cannot be determined, a spatial
variation in polarization angle is equivalent to a change in $RM$. For
the first detection $\Delta RM \approx 110$~\radm, $\Delta RM \approx
40$~\radm\ for the second, and $\Delta RM \approx 8$~\radm\ for the
multi-frequency detection discussed here. If the objects are related,
there are two explanations for the differences in $RM$. First, the
range in $\Delta RM$ could be due to a difference in magnetic field
strength and/or electron density, which would imply that the
contraction of magnetic fields and/or enhancement in electron density
creating the ring must vary over a large range of scales. Secondly, the
range in $\Delta RM$ could be due to a difference in size over which
the RM exhibits a linear gradient with equal $\Delta RM$ per parsec
for each ring. If we assume a distance for the ring detected by Gray
et al.\ to be 1000~pc, then the first two objects would be consistent
with a gradient of $\sim$~3~\radm\ per parsec across the plane of the
sky. If our observed ring would be a feature similar to this, the size
of the ring would be about 3~pc and its distance 60~pc. In this
scenario, rings with similar magnetic field and electron density
characteristics would show a range in size (35~pc, 12~pc and 3~pc).

The first two objects were interpreted to be mainly due to an increase
of electron density. This explanation cannot apply to the ring
discussed here, however, because we observe a change in sign in $RM$
from outside to inside the ring, implying that the cause of the
structure must be at least partly magnetic.

The reason that Gray et al.\ and Uyan\i ker \& Landecker assume that
the rings reflect an increase in electron density are the
following. First, if the structure was magnetic, the magnetic field
enhancement $\Delta B_{\pl}$ would be $\sim$ 5 times the random
component of the magnetic field. This high magnetic field, and a
configuration with the magnetic field directed along the line of sight
was considered to be very unlikely.

However, in their estimate of the $\Delta B_{\pl}$ that is necessary
to produce the observed $\Delta RM$, they assume that the ring-like
structure is an oblate ellipse. If the magnetic field is funnel-shaped
instead, the path length through the structure would increase
enormously and lower $\Delta B_{\pl}$ values can account for the
observed $\Delta RM$. An alignment with the line of sight seems
fortuitous, but it may well be that it is the only configuration that
we can observe, because only such a configuration has a long path
length through a strong $B_{\pl}$, making $RM$s dominant over the
background.
 
We conclude that the magnetic field must play an essential r\^ole in
the shaping of the rings, although an increase in electron density is
likely to also be of importance. The three detected ring-like
structures could be similar objects, although they would need to
exhibit a range in magnetic field and electron density, and/or size.

\section{Structure outside the ring}
\label{s5:out}

\subsection{The elongated structure of high $P$}
\label{ss5:bar}

The elongated structure of high $P$ extending from the center of the
ring to the northwest can be explained with the scenario sketched by Verschuur
(1969). Using the Green Bank 300ft telescope, he observed a
filamentary deficit of \HI, where the filament starts at the position
of the B2V star discussed in Sect.~\ref{s5:intro}, and extends in
the direction opposite to the star's proper motion, along the bar of
high $P$ present in our observations. He argues that the star has
tunneled through the \HI and has blown the neutral material
away. Radiation from the star could then have ionized the remaining
low density material in the tunnel.  This picture is consistent with
the filament of high polarization that we see at the same position,
from the center of the ring to the northwest.

However, the \HI deficit seems to have the same topology as the
ring on the eastern edge, i.e.\ the deficit is shaped as a semi-circle
which coincides with the position of the ring. Then, it is probable
that the ionization tunnel and the ring structure are at same distance
(viz. $\sim$~250~pc), although they do not necessarily have the same
origin.

\subsection{Uniform Galactic magnetic field}

The coherence in $\phi$, and to a lesser extent in $P$, over a large
part of the ring indicates that the polarized foreground and background
have to be uniform on the scale of the ring.  If structure in $RM$ on
smaller scales was present somewhere in the line of sight, be it in
front of or behind the ring-like structure, the linearity of angle
with radius would be destroyed. Small deviations from linearity are
present (see e.g. the patch of less negative $RM$s at $(\alpha,\delta)
= (3^h18^m,66\dg)$), which indicates a change in $RM$ of a few \radm\
somewhere along the line of sight. However, in general the medium must
be very uniform to give such large coherent behavior in the
polarization angle as observed in the ring.

The ring is located in the ``fan region'', which is a region of high
polarization even at low frequencies (Brouw \& Spoelstra, 1976) and
with very ordered polarization angles, located at about $120\dg \la
\ell \la 160\dg$, $0\dg \la b \la 20\dg$. The high polarization in
this region indicates less depolarization, and therefore the uniform
magnetic field should dominate over the random magnetic field
component. From external galaxies, there are indications that indeed
the uniform magnetic field component is higher than the random
component between the spiral arms (Beck \& Hoernes 1996, Beck et al.\
1996).

From Fig.~\ref{f5:rm}, and the $RM$ determinations of Spoelstra (1984)
and Bingham \& Shakeshaft (1967), we derive that outside the ring,
$RM$ $\la 6$~\radm. Then, the parallel component of the magnetic field
must be $B_{\pl} \la 0.2~\mu$G, assuming a path length of the
polarized radiation of 600~pc (Haverkorn et al.\ 2003b) and a thermal
electron density of 0.08~cm$^{-3}$ (Reynolds 1991). The path length is
probably longer than 600~pc, which only reduces $B_{\pl}$ further.
This very low value of $B_{\pl}$ is in agreement with the uniform
magnetic field being mostly perpendicular to the line of sight in this
direction.

\subsection{The aligned depolarization canals}
\label{ss5:can}

\begin{figure}
  \centering
  \psfig{figure=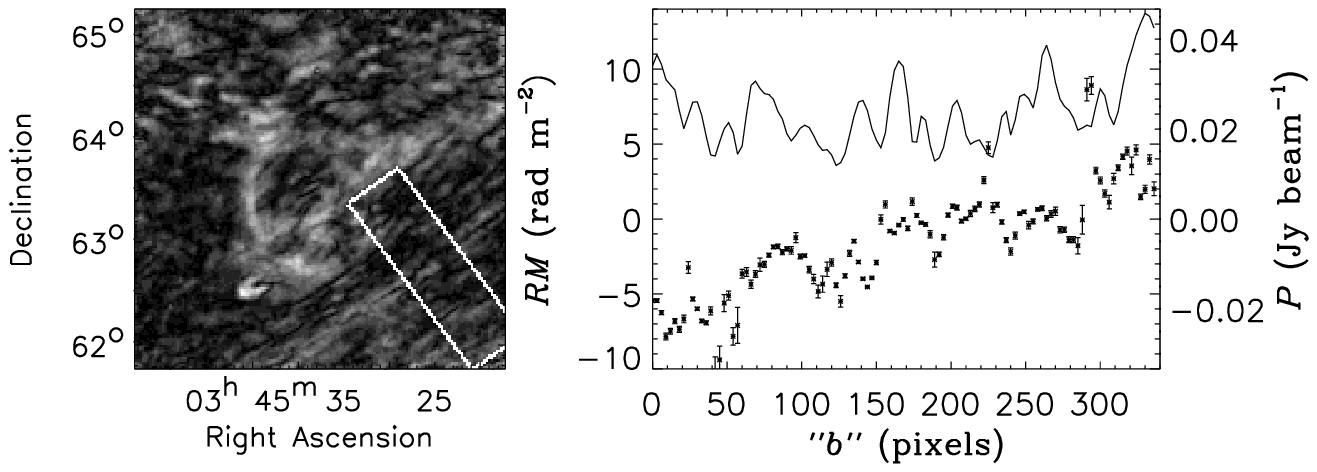,width=.49\textwidth}
  \psfig{figure=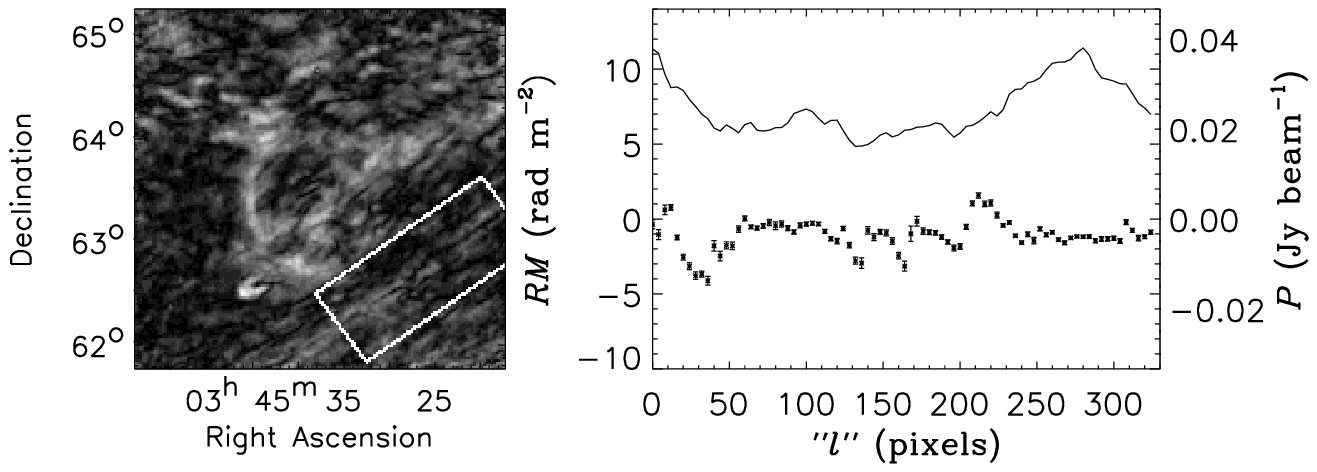,width=.49\textwidth}
  \caption{Left: $P$ at 349~MHz where the white box denotes the
    volume used in the right plot. Right: the symbols denote $RM$
    against a coordinate in the approximate direction of Galactic
    latitude ``$b$'' (top) and longitude ``$\ell$'' (bottom), where
    1~pixel is $\sim$~0.6\arcmin. $RM$ is averaged over ``$\ell$''
    and ``$b$'', respectively. The plotted error in $RM$ is the
    error in the mean of $RM$ in each bin. The upper line is $P$
    at 349~MHz averaged over the same bins.}
  \label{f5:fil}
\end{figure}

In the southwestern part of the field, long and straight
depolarization canals exist that are approximately aligned with
Galactic latitude and one synthesized beam wide, see
Fig.~\ref{f5:pi}. Fig.~\ref{f5:fil} shows the distribution of $RM$ and
$P$ along and across the depolarization 
canals, i.e. approximately along Galactic longitude $\ell$ and
latitude $b$. The top plot shows the $RM$ distribution as a function
of ``$b$'', and the bottom plot as a function of ``$\ell$'', where the
area used is indicated by the white box in the two left-hand
figures. $RM$ and $P$ are averaged in the direction perpendicular to
``$\ell$'' and ``$b$'', respectively. The symbols in the right figures
denote $RM$, where the error bars give the standard error in the mean of $RM$
in each bin. The upper lines denote the $P$ distribution at 349~MHz,
averaged over a bin.  The main difference between the two plots is
that the $RM$ along the filaments hardly changes, whereas the $RM$
across the filaments shows a gradient and/or stratification.

\begin{figure}
  \centering
  \psfig{figure=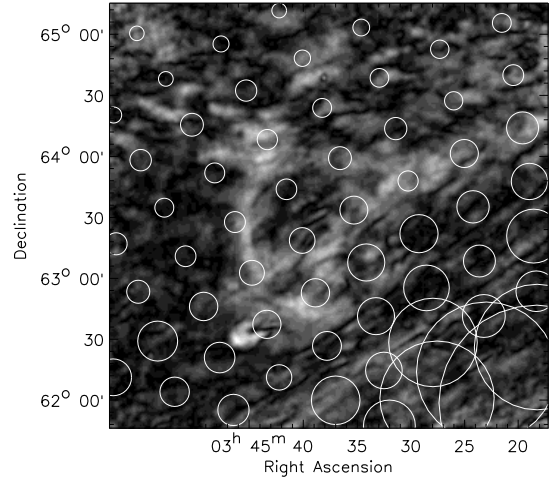,width=.4\textwidth}
  \caption{H$\alpha$ data from the WHAM survey in circles, with a
      resolution of 1\dg, overlaid on polarized intensity at 349~MHz
      in grey scale. The diameters of the circles are proportional to
      the H$\alpha$ intensity, which ranges from $\sim$~2.2~R in the
      upper left corner, to $\sim$~30~R in the lower right corner.}
  \label{f5:wham}
\end{figure}

To investigate whether this $RM$ structure is due to magnetic fields
and/or thermal electron density, we can use independent measurements
of $n_e$, e.g.\ from H$\alpha$ observations. We use data from the
WHAM Northern Sky survey (Haffner et al., in prep., Reynolds et al.\
1998), as shown in Fig.~\ref{f5:wham}. The H$\alpha$ intensity
integrated over the velocity range $v = -10$~km~s$^{-1}$ to $v =
-50$~km~s$^{-1}$ increases from $\sim$~2.2~R to 30~R. This velocity
range coincides with distances of 300~pc to 2.3~kpc using the rotation
curve of Fich et al.\ (1989).
If the H$\alpha$ increase is generated over a line of sight of 2~kpc,
the necessary increase in thermal electron density is only from
0.055~cm$^{-3}$ to 0.07~cm$^{-3}$. If we take again the path length of
the polarized radiation to be 600~pc, the change in $RM$ seen in
Fig.~\ref{f5:fil}, viz 7~\radm, can be made with a constant $B_{\pl}
\approx 0.2~\mu$G, which agrees with the estimate made in
Sect.~\ref{ss5:rm} for the uniform magnetic field component.

Depolarization canals can have two possible causes. The first is
beam depolarization, i.e. the canals denote boundaries between regions
of different $RM$ so that $\Delta\phi = 90\degr$. $RM$ changes with
different magnitudes are likely to exist as well, but these do not
leave visible traces in $P$. The second cause for canals is depth
depolarization. If the emitting and rotating medium is uniform, $P =
\sin(2 RM \lambda^2) / (2 RM \lambda^2)$ so that ``nulls'' of minimum
polarization exist at positions where $2 RM \lambda^2 = n
\pi$. However, there is no reason why depolarization canals would be
one beam wide in this model, and low-$P$ canals only exist for a
uniform medium. Furthermore, the second model predicts the position of
the canals to shift with wavelength, which is not observed. Therefore
we believe that the depolarization canals are predominantly the result
of beam depolarization. For an extended discussion of the causes of
depolarization canals, see Haverkorn et al.\ (2000, 2003a).

\begin{figure}
  \centering
  \psfig{figure=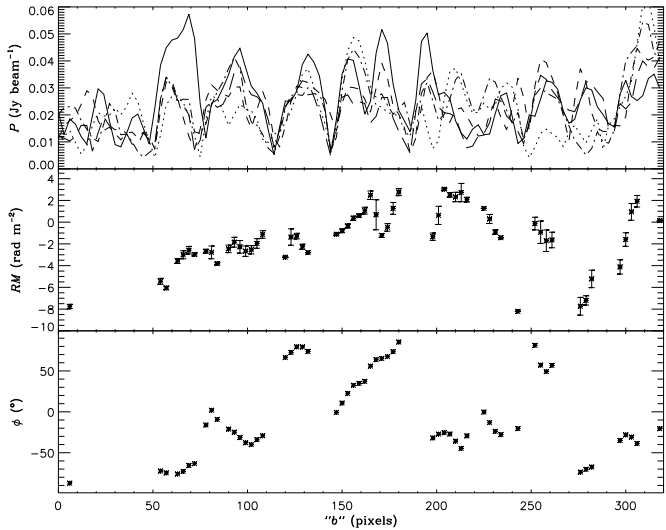,width=.49\textwidth}
  \caption{Spatial variations of $P$ at 5 frequencies (upper plot), $RM$
           (central plot) and $\phi$ (lower plot) over a slit of
           $\sim$~1.5\arcmin\ wide, against ``$b$''. $RM$ and $\phi$
           values are only given where $\chi_{red}^2 < 2$ and
           $\left<P\right> > 20$~mJy/beam.}
  \label{f5:fil_n}
\end{figure}

If the gradient in $RM$ is constant, depolarization canals due to
beam depolarization would not be possible. So the $RM$ must
increase discontinuously, as is visible in Fig.~\ref{f5:fil}. If
$\Delta RM$ over one beam is between 2.1 and 2.5~\radm\ (or between
6.3 and 7.5~\radm, etc.), depolarization canals occur.
Fig.~\ref{f5:fil_n} shows $P$ and $RM$ over a narrow slit
($\sim$~1.5\arcmin) across the canals plotted against $b$, located
within the box in Fig.~\ref{f5:fil}. The top panel shows plots for $P$
at all 5 frequencies, oversampled by a factor of 5. The central panel
shows $RM$, where $RM$ values are only plotted for beams for which
$\chi_{red}^2 < 2$ and $\left<P\right> > 20$~mJy/beam. Polarization
angle $\phi$ is given in the bottom panel. The two sharpest
depolarization canals visible at all frequencies, at pixel numbers 113
and 145 correspond to abrupt changes in $RM$ of about 2~\radm\ and
angle changes around 90\dg. Abrupt $RM$ changes with other magnitudes
are present as well, but these do not create depolarization canals. 

\section{Conclusions}
\label{s5:conc}

The ring-like structure in polarized intensity~$P$ with a radius of
about 1.4\dg\ shows a regular increase in polarization angle from its
center out to $\sim$~1.7\dg, suggesting that the structure is a disk
instead of a ring, and extends to larger radii than the ring in
$P$. The rotation measure is slightly positive outside the ring,
reverses sign inside the ring and decreases almost continuously to
$RM~\approx -8$~\radm\ at the center. This property rules out its
production by any spherically symmetric structure, such as a supernova
remnant or a wind-blown bubble. The $RM$ structure, and the
observation that the coherence in angle slowly disappears beyond a
radius of $\sim$~1.7\dg\ indicates a magnetic origin for the
polarization ring, probably accompanied by an electron density
enhancement. We propose that the ring is produced by a predominantly
magnetic funnel-like structure, in which the parallel magnetic field
strength is maximal at the center and directed 
away from us.  The enhancement in $P$ at radius $\sim$~1.4\dg\ is
caused by a lack of depolarization due to the relative constancy of
$RM$ at that radius. A filamentary pattern of parallel, narrow
depolarization canals indicates structure in $RM$ which is aligned
with Galactic longitude. The depolarization canals are probably
created by beam depolarization due to abrupt spatial gradients in $RM$.

\begin{acknowledgements}
We are grateful to T. Spoelstra for allowing us to use his
observations, and to R. Beck, E. Berkhuijsen and the anonymous referee
for suggestions and comments on the manuscript. The Westerbork Synthesis
Radio Telescope is operated by the Netherlands Foundation for Research
in Astronomy (ASTRON) with financial support from the Netherlands
Organization for Scientific Research (NWO). The Wisconsin H-Alpha
Mapper is funded by the National Science Foundation. MH is supported
by NWO grant 614-21-006.
\end{acknowledgements}

\end{document}